\documentclass[aps,pra,reprint,showpacs,twocolumn,superscriptaddress,eqsecnum]{revtex4-1}
\usepackage{graphicx}
\usepackage{amsmath}
\usepackage{amssymb}
\usepackage{amsfonts}
\usepackage{bbm}
\usepackage{bm}
\usepackage{verbatim}
\usepackage{cancel}
\usepackage{tikz}
\usepackage{wasysym}
\usepackage[bookmarks=false,colorlinks=true,urlcolor=blue,citecolor=blue,linkcolor=blue]{hyperref}
\usepackage[normalem]{ulem}
\allowdisplaybreaks

\tikzstyle{tensor}=[rectangle,draw=blue!50,fill=blue!20,thick]

\begin{document}

\title{Quantum Many-Body Scar States with Emergent Kinetic Constraints and Finite-Entanglement Revivals}

\author{Thomas Iadecola}
\affiliation{Condensed Matter Theory Center and Joint Quantum Institute, Department of Physics, University of Maryland, College Park, Maryland 20742, USA}
\affiliation{Department of Physics and Astronomy, Iowa State University, Ames, Iowa 50011, USA}

\author{Michael Schecter}
\affiliation{Condensed Matter Theory Center and Joint Quantum Institute, Department of Physics, University of Maryland, College Park, Maryland 20742, USA}

\date{\today}

\begin{abstract}
We construct a set of exact, highly excited eigenstates for a nonintegrable spin-1/2 model in one dimension that is relevant to experiments on Rydberg atoms in the antiblockade regime.  These states provide a new solvable example of quantum many-body scars: their sub-volume-law entanglement and equal energy spacing allow for infinitely long-lived coherent oscillations of local observables following a suitable quantum quench. While previous works on scars have interpreted such oscillations in terms of the precession of an emergent macroscopic SU(2) spin, the present model evades this description due to a set of emergent kinetic constraints in the scarred eigenstates that are absent in the underlying Hamiltonian. We also analyze the set of initial states that give rise to periodic revivals, which persist as approximate revivals on a finite timescale when the underlying model is perturbed.  Remarkably, a subset of these initial states coincides with the family of area-law entangled Rokhsar-Kivelson states shown by Lesanovsky to be exact ground states for a class of models relevant to experiments on Rydberg-blockaded atomic lattices.
\end{abstract}

\maketitle

\section{Introduction}

The ergodic hypothesis holds that a generic quantum many-body system prepared in a simple initial state relaxes under unitary evolution to a steady state that depends only on the initial values of conserved quantities, such as energy or particle number.  For the present purposes, a ``generic" quantum system is one that is strongly interacting and far from any integrable regions of parameter space, and a ``simple" initial state is one that can be prepared by minimizing the energy of a local Hamiltonian.  While this point of view is reasonable, and essentially unavoidable in systems where the eigenstate thermalization hypothesis (ETH) holds~\cite{Deutsch91,Srednicki94,Rigol08,D'Alessio16,Deutsch18}, a growing body of recent work has begun to grapple with systems where it does not apply~\cite{Nandkishore15,Abanin19}.

One striking development along these lines has been the emergence of quantum many-body scars, which stems from a surprising experimental observation~\cite{Bernien17} in a Rydberg-blockaded atomic chain of length $L=51$: when the system was prepared in a N\'eel state and allowed to evolve unitarily, it was found that measurements of local observables exhibited long-lived coherent oscillations. Such behavior is unexpected given the finite energy density of the initial state and the nonintegrable nature of the relevant model Hamiltonian~\cite{Rigol08}.  These oscillations were later attributed to the existence of a set of rare eigenstates (of order $L$ in number) having large overlap with the N\'eel state~\cite{Turner17}.  Furthermore, these ``scarred" eigenstates are (nearly) equally spaced in energy and, according to finite-size numerics, appear to violate the entanglement ``volume law" by exhibiting entanglement entropy scaling as $\ln L$~\cite{Turner18}.  These three anomalous features, namely
\begin{align*}
\noindent 
&\text{$i)$ concentration of weight on a ``simple" initial state,}\\
&\text{$ii)$ (nearly) equal energy spacing, and}\\
&\text{$iii)$ sub-volume-law entanglement,}
\end{align*}
enable long-lived coherent oscillations of local observables in an otherwise thermalizing system and can be taken as an operational definition of scarred eigenstates.  QMBS thus provide an intriguing example of \textit{weak} ergodicity breaking, where the preparation of a \textit{particular} initial state can lead to nonthermalizing quantum dynamics~\cite{Biroli10,Deutsch18}.

The initial discovery of QMBS brought into focus a number of pressing questions, foremost among them being the microscopic origin of the scarred many-body states in the experimentally relevant model, known colloquially as the ``PXP model." A number of possibilities have been put forward, including various quasiparticle pictures~\cite{Lin18,Iadecola19}, proximity to a putative integrable point~\cite{Khemani18}, connections to lattice gauge theories~\cite{Surace19,Magnifico19}, and the potential existence of an emergent macroscopic SU(2) ``spin" embedded in the many-body spectrum~\cite{Choi18}. Reconciling these diverse (though not necessarily contradictory) viewpoints is challenging given the relatively poor analytical understanding of the PXP model, which necessitates a reliance on finite-size numerics.

An alternative approach is to find analytically tractable nonintegrable models with scarred eigenstates and explore their properties in hopes of developing a useful framework for comparison with experiments.  Exact finite-energy-density eigenstates with equal energy spacing and sub-volume-law entanglement [i.e., satisfying conditions $ii)$ and $iii)$] were found in the Affleck-Kennedy-Lieb-Tasaki (AKLT) spin chain in Refs.~\cite{Moudgalya18a,Moudgalya18b}. However, it is not yet known what quench is necessary to observe revivals of the many-body wavefunction---i.e., it is not known whether condition $i)$ is satisfied in this model.  In Ref.~\cite{Schecter19}, we demonstrated that another model, the spin-1 $XY$ model, also possesses a set of atypical eigenstates satisfying $ii)$ and $iii)$, and furthermore that condition $i)$ holds for these eigenstates.  The necessary low-entanglement initial state (in this case a product state) was identified owing to the existence of an emergent SU(2) algebra similar to the one suggested on phenomenological grounds in Ref.~\cite{Choi18}: in this model, the revivals of the many-body wavefunction following a quench from this initial state are interpreted in terms of a macroscopic SU(2) spin precessing in a magnetic field.

Here we show that this simple SU(2) picture is not the whole story by constructing a new exact instance of QMBS in a class of spin-1/2 models.  Like the states constructed in Refs.~\cite{Moudgalya18a,Moudgalya18b,Schecter19}, the scars in these models form towers of states containing ``condensates" of stable quasiparticles at momentum $\pi$.  However, unlike the ones studied in Ref.~\cite{Schecter19}, these quasiparticles cannot be annihilated by a local operator when a finite density of them are present.  This is due to the emergence of a kinetic constraint that forbids them from occupying neighboring sites (a similar constraint was found in Refs.~\cite{Moudgalya18a,Moudgalya18b}).  This kinetic constraint is precisely the ``Fibonacci" constraint that arises in the experimentally relevant Rydberg-blockaded model~\cite{Lukin01,Lesanovsky12b}, but in this case is emergent in that it is not present in the underlying Hamiltonian and its other eigenstates.  Thus, the ``raising" and ``lowering" operators for the towers of states we construct cannot be adjoints of one another. This explicitly precludes the grouping of scarred eigenstates into representations of an SU(2) algebra.  

Despite this, we find a set of initial states giving rise to exact revivals. This set includes a family of initial states that, unlike ones found in previous exact examples of QMBS, are not product states but finitely-entangled area-law states.  Remarkably, this family of initial states is related to a set of Rokhsar-Kivelson states shown by Lesanovsky~\cite{Lesanovsky11} to be the exact ground states of a class of models relevant to experiments in Rydberg-blockaded atomic lattices.  While the possibility of periodic revivals to a state with finite entanglement was raised in Ref.~\cite{Michailidis19} in the context of a time-dependent variational principle (TDVP) study, the present work demonstrates that such behavior can arise in an analytically exact setting.  We also demonstrate that the exact periodic revivals give way to highly coherent approximate revivals on a finite timescale, similar to what is seen in the PXP model~\cite{Turner17} and other models with approximate scar states~\cite{Moudgalya19,Bull19,Hudomal19}, when the model is perturbed.

We further find a prescription based on an auxiliary SU(2) algebra for choosing initial states that exhibit revivals. Despite the fact that the scarred eigenstates do not form a representation of this algebra, it is possible to project other states that \textit{do} form an appropriate representation into the kinetically constrained subspace to obtain the desired initial states. The resulting dynamics still does not have an SU(2) structure as the kinetic constraint selects a preferred quantization axis.

The paper is organized as follows. In Sec.~\ref{sec: Model and Scarred Eigenstates} we define the model and show that it possesses two towers of scarred eigenstates, related by a $\mathbb Z_2$ symmetry. We exemplify their nonthermal nature both by an exact calculation of their entanglement spectrum and by finding a local Hamiltonian whose ground state manifold is spanned by these states.  In Sec.~\ref{sec: Structure of the Scarred Eigenstate Towers} we examine the structure of the scarred eigenstate towers and show how the emergent kinetic constraint 
precludes the scarred eigenstates from forming a representation of an SU(2) algebra.  In Sec.~\ref{sec: Initial States and Finite-Entanglement Revivals}, we identify the relevant initial states and consider the effect of perturbations. Finally, we offer conclusions in Sec.~\ref{sec: Conclusions}.

\section{Model and Scarred Eigenstates}
\label{sec: Model and Scarred Eigenstates}

We consider spin-1/2 degrees of freedom on a chain with $L$ sites described by the Hamiltonian (see also Ref.~\cite{Ostmann19})
\begin{align}
\label{eq: H}
\begin{split}
H\!&=\!\sum^{i_L}_{i=i_1}\!\left[\lambda\, (\sigma^x_i-\sigma^z_{i-1}\sigma^x_{i}\sigma^z_{i+1})+\Delta\,\sigma^z_i+J\, \sigma^z_{i}\sigma^z_{i+1}\right]\\
&\equiv H_{\lambda}+H_{z}+H_{zz}
\end{split}
\end{align}
where $\sigma^{x,y,z}_i$ are Pauli matrices defined on site $i$. We hereafter set $\lambda=1$ and, without loss of generality, choose $L$ to be even.  The limits on the sum above depend on whether open or periodic boundary conditions (OBC or PBC) are chosen.  For OBC, we take $i_1=2$ and $i_L=L-1$, so that $[H,\sigma^z_{1,L}]=0$.  In this case, the edge spins are constants of motion and can be fixed from the outset.  For PBC, $i_1=1$ and $i_L=L$, with sites $0\equiv L$ and $L+1 \equiv 1$.  Regardless of boundary conditions, the model has a spatial inversion symmetry $I$ with eigenvalues $\pm 1$ and a $U(1)$ symmetry associated with conservation of the number of Ising domain walls, $n_{\rm DW}$.

It is useful to consider how the Hamiltonian \eqref{eq: H} acts on $z$-basis product states, which we represent in terms of the local basis
\begin{align}
\sigma^z_i|1\rangle=|1\rangle,\indent\sigma^z_i|0\rangle=-|0\rangle.
\end{align}
In this case, the only off-diagonal part of $H$ is $H_\lambda$, which flips a spin if and only if its nearest neighbors are in different spin states; the effect of this dynamical rule is to enforce conservation of $n_{\rm DW}$.  For example, up to a prefactor, $H_\lambda$ maps
\begin{align}
\label{eq: scattering}
|\dots00100\dots\rangle\!\to\!|\dots01100\dots\rangle\!+\!|\dots00110\dots\rangle,
\end{align}
which manifestly conserves the Ising domain-wall number.

\begin{figure}[t!]
\begin{center}
\includegraphics[width=.9\columnwidth]{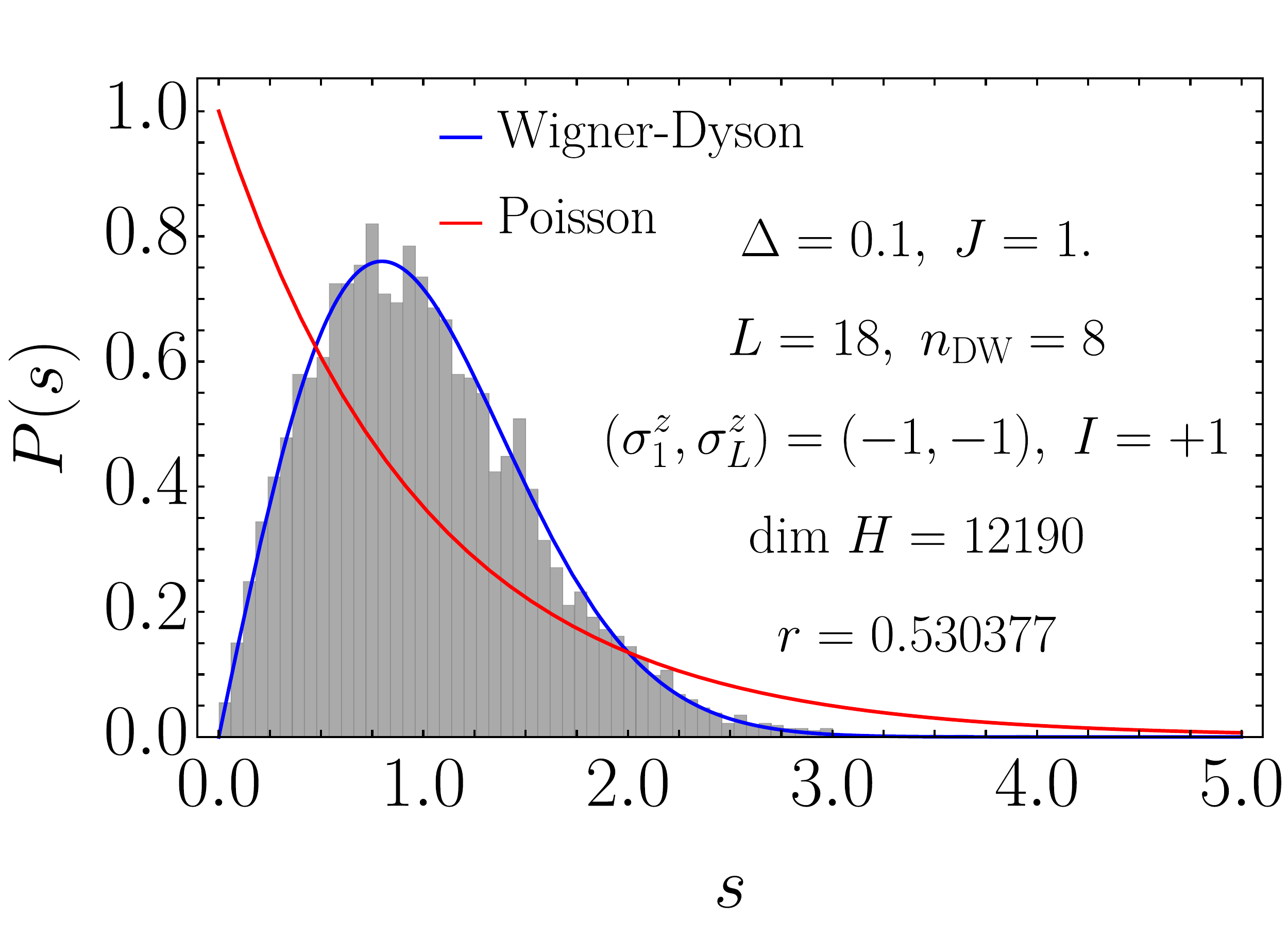}
\caption{Exact diagonalization results on statistics of energy level spacings in the middle half of the spectrum for the model \eqref{eq: H} with OBC.  The Hamiltonian parameters and symmetry sectors used are indicated in the figure.  The Poisson (red) and Wigner-Dyson (blue) distributions, the former characteristic of integrable systems and the latter of chaotic systems described by random matrix theory, are shown for comparison.  Excellent agreement with the Wigner-Dyson distribution is shown as, quantified by the $r$-value of the distribution~\cite{Pal10}, which is very close to the Wigner-Dyson value $r_{\rm WD}\approx 0.5295$.}
\label{fig: lvl}
\end{center}
\end{figure}

For generic values of $\Delta$ and $J$, the model \eqref{eq: H} is nonintegrable.  This can be verified, e.g., by examining the probability distribution of the spacings $s$ between energy levels, $P(s)$, which appears to be well described by random matrix theory for levels in a given symmetry sector (see Fig.~\ref{fig: lvl}).  When $\Delta=0$, the model is integrable and can be solved by mapping to free fermions using a global spin rotation followed by a Jordan-Wigner transformation.  Adding a finite $\Delta$ introduces a highly nonlocal interaction term in the fermionic language, breaking integrability.

\subsection{Definition of Scarred States}
\label{sec: Definition of Scarred States}

We are now in a position to define the scarred eigenstates of the model \eqref{eq: H}.  We first define the tower of states
\begin{subequations}
\label{eq: Sn}
\begin{align}
\label{eq: Sn a}
|\mathcal S_n\rangle=\frac{1}{n!\sqrt{\mathcal N(L,n)}}\,(Q^\dagger)^n\, |\Omega\rangle,
\end{align}
where $|\Omega\rangle=|0\dots0\rangle$ and
\begin{align}
\mathcal N(L,n)
=
\begin{cases}
\binom{L-n-1}{n} & \text{OBC}\\
\frac{L}{n}\binom{L-n-1}{n-1} & \text{PBC}\\
\end{cases}.
\end{align}
The operator
\begin{align}
\label{eq: Sn b}
Q^\dagger=\sum^{i_L}_{i=i_1}(-1)^i\, P^0_{i-1}\sigma^{+}_{i}P^0_{i+1},
\end{align}
where $\sigma^\pm_j=(\sigma^x_j \pm i\,\sigma^y_j)/2$ and
$P^0_i = (1-\sigma^z_i)/2$
is the local projector onto spin down.
When PBC are imposed, the total momentum of the state $|\mathcal S_n\rangle$ is equal to $K_n=\pi n\ (\text{mod }2\pi)$; for OBC and PBC, its inversion quantum number is $(-1)^n$ and its $U(1)$ quantum number is $n_{\rm DW}=2n$.
\end{subequations}
The state $|\mathcal S_n\rangle$ has energy
\begin{align}
\label{eq: En}
E_n=(2\Delta-4J)\,n+J(L-1)-\Delta L.
\end{align}
Note that for generic $n$ of order $L$, the states $|\mathcal S_n\rangle$ have finite energy density, corresponding to infinite temperature for generic parameters in Eq.~\eqref{eq: H}. The proof that the states \eqref{eq: Sn} are eigenstates of Eq.~\eqref{eq: H} relies on showing that $H_{\lambda}|\mathcal S_n\rangle=0$ and proceeds along the lines of the analogous calculation in Ref.~\cite{Moudgalya18a}; we present it in Appendix~\ref{sec: eigenstate proof}.

Physically, the state $|\mathcal S_n\rangle$ contains $n$ magnons (i.e., spin flips, or 1s in a background of 0s), each carrying momentum $k=\pi$.  In fact, when $n/L$ is finite it can be viewed as a condensate of such magnons, as it possesses off-diagonal long-range order (ODLRO)~\cite{Yang62} with respect to the ``order parameter" $Q^\dagger$, similar to Refs.~\cite{Yang89,Schecter19}. Physically, this ODLRO manifests itself in long-range \textit{connected} magnetic correlations.  This fact itself is evidence that these states generically do not obey the ETH; an infinite-temperature state has a trivial density matrix and therefore cannot support long-range connected correlations.

The magnons in these states are subject to a kinetic constraint: two magnons cannot occupy neighboring sites owing to the projectors $P^0_i$ in $Q^\dagger$.  We call this the ``Fibonacci constraint" as the number of states in the spin-1/2 Hilbert space satisfying this constraint grows as a power of the golden ratio~\cite{Lesanovsky12b}.  While this kinetic constraint arises naturally in Rydberg-blockaded atomic lattices and is built into the PXP model, in the present context it is \textit{emergent} in the sense that this constraint is not present, even approximately, in the underlying Hamiltonian~\eqref{eq: H}.  By considering the form of the states \eqref{eq: Sn} in the case where the operator $Q$ does \textit{not} contain the projectors $P^0$, one can see that this kinetic constraint is necessary in order to ensure that the state $|\mathcal S_n\rangle$ is an eigenstate of $n_{\rm DW}$.  For instance, the configuration $|\dots00110\dots\rangle$ contains the same number of magnons as $|\dots01010\dots\rangle$, but has a different $n_{\rm DW}$ eigenvalue. The emergence of the constraint is therefore tied to the U(1) symmetry of the model \eqref{eq: H}.

The kinetic constraint implies that the number of states $|\mathcal S_n\rangle$ depends on boundary conditions. For PBC, there are $L/2+1$ such states ($n=0,\dots,L/2$), since $(Q^\dagger)^{L/2+1}=0$ owing to the kinetic constraint.  For OBC, there are only $L/2$ ($n=0,\dots,L/2-1$), since the edge spins are frozen [see discussion below Eq.~\eqref{eq: H}], effectively reducing $L$ by $2$.

In addition to the tower of states defined in Eqs.~\eqref{eq: Sn}, there is another tower of exact eigenstates related to the states $|\mathcal S_n\rangle$, namely
\begin{subequations}
\label{eq: Snp}
\begin{align}
\label{eq: Snp a}
|\mathcal S^{\prime}_n\rangle = G\, |\mathcal S_n\rangle=\frac{1}{n!\sqrt{\mathcal N(L,n)}}\,(Q^{\prime\,\dagger})^n\, |\Omega^\prime\rangle,
\end{align}
where $G=\prod^L_{i=1}\sigma^x_i$ is a $\mathbb Z_2$ transformation that flips all spins, $|\Omega^\prime\rangle=|1\dots1\rangle$, and
\begin{align}
\label{eq: Snp b}
Q^{\prime\, \dagger}=G\, Q^\dagger\,G=\sum^{i_L}_{i=i_1}(-1)^i\, P^1_{i-1}\sigma^{-}_{i}P^1_{i+1},
\end{align}
\end{subequations}
with
$P^1_i = (1+\sigma^z_i)/2$
the local projector onto spin up.
The energy of the state $|\mathcal S^{\prime}_n\rangle$ is given by
\begin{align}
E^\prime_n=-(2\Delta+4J)\,n+J(L-1)+\Delta L,
\end{align}
and the remaining symmetry quantum numbers are identical to the states $|\mathcal S_n\rangle$.

The states $|\mathcal S^{\prime}_n\rangle$ have the same interpretation as their counterparts $|\mathcal S_n\rangle$, except that the roles of the 0 and 1 states are interchanged.  The dependence of the number of states $|\mathcal S_n^\prime\rangle$ is identical to that of $|\mathcal S_n\rangle$.  Taking both towers of states into account, there are $L$ scarred eigenstates with OBC, while there are $L+1$ scarred eigenstates with PBC.  In the latter case, the extra state comes from the N\'eel cat state $(|1010\dots\rangle\pm|0101\dots\rangle)/\sqrt{2}$, which contains the maximum possible number of magnons given the kinetic constraint (the $\pm$ depends on $L$ and must be chosen such that the state has the appropriate momentum quantum number).  Since this state is $\mathbb Z_2$-symmetric, it belongs to both towers: $|\mathcal S_{L/2}\rangle = |\mathcal S^\prime_{L/2}\rangle$ for PBC~\footnote{Note that, strictly speaking, both N\'eel cat states are eigenstates of Eq.~\eqref{eq: H} regardless of boundary conditions.  What depends on $L$ and the choice of boundary conditions is whether these states can be reached by successive application of the operators \eqref{eq: Sn b} and \eqref{eq: Snp b}.}.  As these two towers of states are related by applying a simple unitary operator, we will hereafter consider only the tower of states $|\mathcal S_n\rangle$ with the understanding that similar results hold for the states $|\mathcal S^\prime_n\rangle$.  We will also restrict to OBC, with the understanding that all results for OBC have analogs for PBC.

\subsection{Sub-Volume-Law Entanglement}
\label{sec: Sub-Volume-Law Entanglement}

\begin{figure}[t!]
\begin{center}
\includegraphics[width=.87\columnwidth]{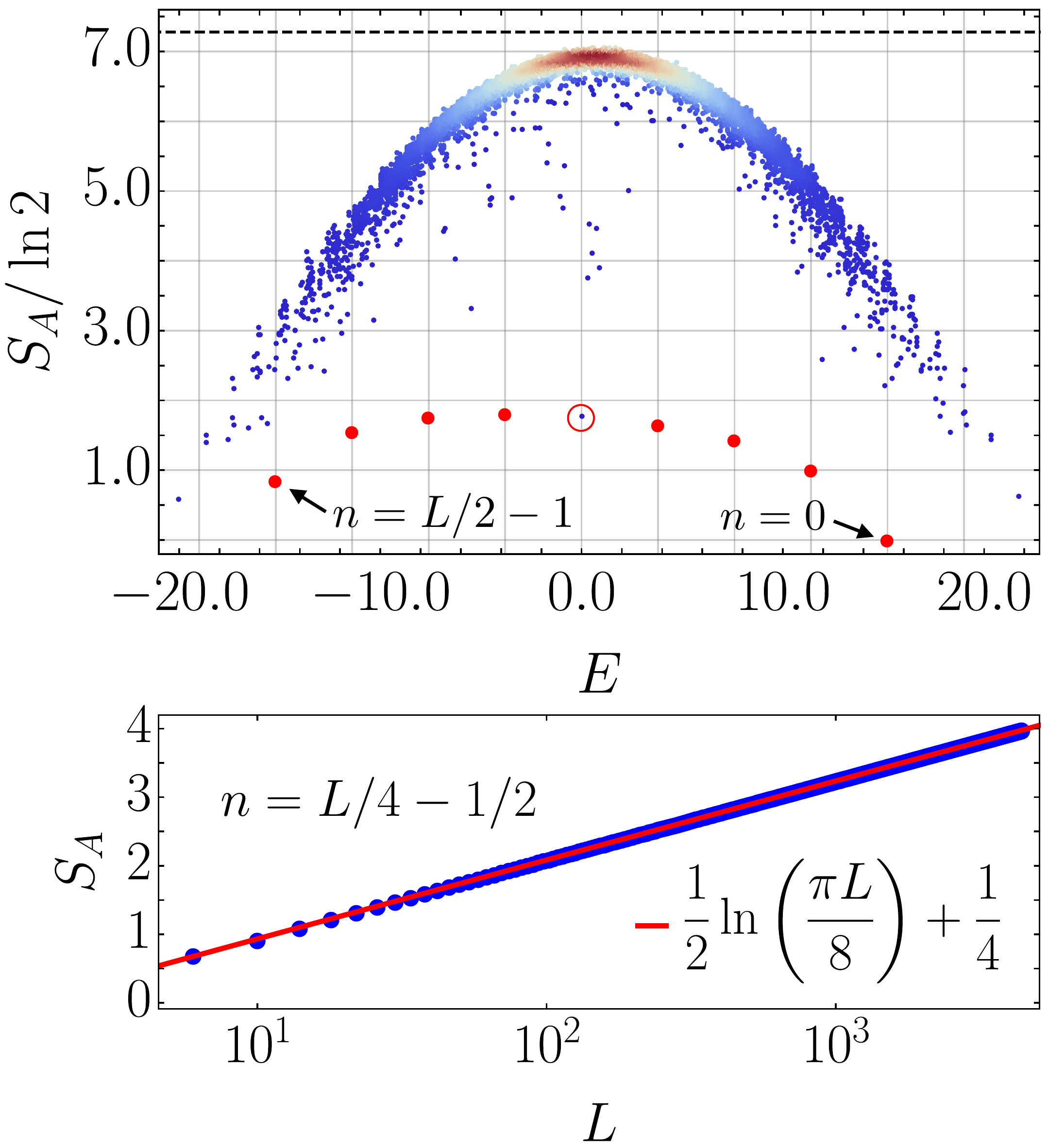}\\
\caption{
Bipartite entanglement entropy $S_A$ [Eq.~\eqref{eq: S_A}] in the model \eqref{eq: H}; $L=18$ and Hamiltonian parameters are the same as in Fig.~\ref{fig: lvl}.
Top: $S_A$ vs.~many-body energy $E$.  Eigenstates in the symmetry sector with $n_{\rm DW}=8$, where the scarred state $|\mathcal S_{n=4}\rangle$ (circled) resides, are represented by small points.  The points are color coded
according to their density, such that warmer colors indicate more densely packed points.  Larger red points indicate the analytical values of $S_A$ for the scarred states $|\mathcal S_{n\neq 4}\rangle$, which reside in symmetry sectors with different values of $n_{\rm DW}$ and $I$.  The dashed black line indicates the expected value for a random state, Eq.~\eqref{eq: S_A ran}. Bottom: Exact $S_A(L,L/2,n)$ [Eq.~\eqref{eq: S_A exact}] for the scarred state $|\mathcal S_n\rangle$ at $n=L/4-1/2$ for a family of system sizes $L=2\ \text{mod}\ 4$, which includes $L=18$.  Data are plotted on a log-linear scale, with a line of best fit drawn in red.
}
\label{fig: ee}
\end{center}
\end{figure}

We now demonstrate that the states defined in Eqs.~\eqref{eq: Sn} generically have sub-volume-law entanglement, thereby demonstrating explicitly their ETH-violating nature.  We quantify the entanglement of a state $|\psi\rangle$ using the von Neumann entanglement entropy, defined with respect to a bipartition of the system into subsystems $A$ and $B$ as
\begin{align}
\label{eq: S_A}
S_A=-\text{tr}(\rho_A\ln\rho_A)
\end{align}
where $\rho_A = \text{tr}_{B}|\psi\rangle\langle\psi|$ is the reduced density matrix on region $A$, which we take to be the left half of the system.  Numerical results from exact diagonalization at $L=18$ are shown in the top panel of Fig.~\ref{fig: ee}.  These data clearly show that typical eigenstates in the middle of the many-body spectrum have entanglement entropy close to the expected value for a random state~\cite{Page93},
\begin{align}
\label{eq: S_A ran}
S^\mathrm{ran}_A=\frac{L-2}{2}\ln 2-\frac{1}{2},
\end{align}
indicated by the dashed black line in the top panel of Fig.~\ref{fig: ee}. (Here $L-2$ appears rather than $L$ due to the freezing of the edge spins under OBC.) This value scales with $L$, exemplifying the entanglement volume law expected in high-energy-density eigenstates of generic local Hamiltonians.  The scarred eigenstates (larger red points in Fig.~\ref{fig: ee}), on the other hand, have much lower entanglement than typical eigenstates nearby in energy.  While these numerical results are for fixed $L$, an exact calculation of the entanglement spectrum (i.e., the eigenvalues of $\rho_A$) allows for the evaluation of $S_A$ in the states $|\mathcal S_n\rangle$ for arbitrary system sizes, where one finds $S_A\sim\ln L$ for generic $n$ (see bottom panel of Fig.~\ref{fig: ee}).

We compute the entanglement spectrum for a state $|\mathcal S_n\rangle$ by writing
\begin{align}
\label{eq: Sn cut}
\begin{split}
|\mathcal S_n\rangle 
&= \frac{1}{\sqrt{\mathcal N(L,n)}}
\sum_{\substack{i_1<\dots<i_n\\ |i_p-i_{p+1}|>1\\ i_p\neq 1,L}}
\!\!\!
\sigma(\{i_p\})\,
|\{i_p\}\rangle\\
&\equiv
\sum_{\substack{i_1<\dots<i_n\\ |i_p-i_{p+1}|>1\\ i_p\neq 1,L}}
M_{\{i_p\}_A,\{i_p\}_B}\,
|\{i_p\}_A\rangle\otimes|\{i_p\}_B\rangle.
\end{split}
\end{align}
Here, we denote by $|\{i_p\}\rangle$ a state with $n$ magnons at positions $i_p$, $p=1,\dots, n$, and $\sigma(\{i_p\})=(-1)^{\sum^n_{p=1}i_p}$.  The set $\{i_p\}_{A(B)}$ denotes the restriction of the set $\{i_p\}$ to region $A$ ($B$), which we take to have length $L_{A}$ ($L_{B}$).  The entanglement spectrum is then given by the eigenvalues of the matrix $\mathcal M \equiv MM^\dagger$.  In Appendix~\ref{sec: es}, we show that $\mathcal M = \bigoplus^{K}_{k=0}\mathcal M_{k}$ [$K=\text{min}(n,\lfloor L_A/2 \rfloor)$ for OBC], where the subblocks $\mathcal M_{k}$ each contribute a pair of eigenvalues
\begin{subequations}
\label{eq: es}
\begin{widetext}
\begin{align}
\label{eq: lambda}
\lambda_{k,\pm}(L,L_A,n)=\frac{D_{1,k}m_{1,k}+D_{2,k}m_{2,k}\pm\sqrt{(D_{1,k} m_{1,k} - D_{2,k} m_{2,k})^2 + 4 D_{1,k} D_{2,k} m_{2,k}^2}}{2\mathcal N},
\end{align}
\end{widetext}
where $\mathcal N\equiv \mathcal N(L,n)$ and
\begin{align}
\label{d1k}
D_{1,k}\equiv D_{1,k}(L_A) &= \mathcal N(L_A,k)  \\ 
\label{d2k}
D_{2,k}\equiv D_{2,k}(L_A) &= \mathcal N(L_A-1,k-1)  \\
\label{m1k}
m_{1,k}\equiv m_{1,k}(L_B,n) &= \mathcal N(L_B,n-k) \\
&\qquad+\mathcal N(L_B-1,n-k-1) \nonumber
\\
\label{m2k}
m_{2,k}\equiv m_{2,k}(L_B,n) &= \mathcal N(L_B,n-k). 
\end{align}
\end{subequations}
The exact entanglement entropy for the scar state $|\mathcal S_n\rangle$ is then
\begin{align}
\label{eq: S_A exact}
S_A(L,L_A,n)=-\sum^n_{k=0} \sum_{s=\pm}\lambda_{k,s}\ln\lambda_{k,s}.
\end{align}
This analytical expression agrees to numerical precision with the values of $S_A$ extracted from exact diagonalization (e.g., in the top panel of Fig.~\ref{fig: ee}).
A plot of Eq.~\eqref{eq: S_A exact} as a function of $L$ with $L=2\ \text{mod}\ 4$, $L_A=L/2$, and $n=L/4-1/2$ is shown in the bottom panel of Fig.~\ref{fig: ee}.  A clear logarithmic scaling with $L$ is observed.  This scaling is consistent with results found for the exact eigenstates in Refs.~\cite{Moudgalya18b,Schecter19}, which also have an interpretation in terms of free or kinetically constrained quasiparticles. 

\subsection{Scarred Eigenstates as Ground States}
\label{sec: Characteristic Projectors}

Another remarkable feature of the states $|\mathcal S_n\rangle$ is that there exists a local Hamiltonian whose ground state manifold is spanned by these states. This is yet another feature of the states $|\mathcal S_n\rangle$ that contradicts expectations for ETH-obeying states. For example, while ground states of generic local Hamiltonians in 1D are believed to violate the area law at most logarithmically~\cite{Eisert10} (see though~\cite{Movassagh16} and references therein), ETH-obeying states are always volume-law entangled.  The discussion below further highlights the local structure of the states $|\mathcal S_n\rangle$.

The aforementioned local Hamiltonian is given by
\begin{subequations}
\begin{align}
\label{eq: Hu}
H_{\rm u} = \sum^{i_L}_{i=i_1}\mathcal P_i,
\end{align}
with
\begin{align}
\label{eq: Pi def}
\mathcal P_i = P^1_{i}\, P^1_{i+1} + P^0_{i-1}\left(\frac{1}{4}+\bm S_{i} \cdot \bm S_{i+1}\right)P^0_{i+2},
\end{align}
where we have defined
\begin{align}
\bm S_{i} = \frac{1}{2}\, \left(\sigma^x_i,\sigma^y_i,(-1)^i\sigma^z_{i}\right)^\mathsf{T}.
\end{align}
\end{subequations}
In Appendix~\ref{sec: uncle}, we show using matrix product state (MPS) techniques that
\begin{align}
\mathcal P_i\, |\mathcal S_n\rangle = 0 \qquad \forall\ i,n,
\end{align}
i.e., that the states $|\mathcal S_n\rangle$ are frustration-free ground states of the Hamiltonian $H_{\rm u}$.  As might be anticipated from the $\ln L$ scaling of entanglement in the states $|\mathcal S_n\rangle$, finite-size numerics suggest that the Hamiltonian \eqref{eq: Hu} is gapless in the thermodynamic limit.  For this reason, the Hamiltonian $H_{\rm u}$ can be viewed as an ``uncle Hamiltonian"~\cite{Fernandez15} for the states $|\mathcal S_n\rangle$, the term ``parent Hamiltonian" being reserved for gapped Hamiltonians having a particular MPS as their ground state.

Having identified the projectors $\mathcal P_i$, it is now possible to write down another class of local Hamiltonians where the states $|\mathcal S_n\rangle$ appear as ``scarred" many-body eigenstates at finite energy density.  This is a class of ``embedded" Hamiltonians~\cite{Shiraishi17,Mondaini18,Shiraishi18,Ok19} of the form
\begin{align}
\label{eq: He}
H_{\rm e} = H_0+H'\equiv \sum_{i}\mathcal P_i\, h_i\, \mathcal P_i + H',
\end{align}
where $h_i$ is any local interaction acting within a few sites of site $i$, and where $H'$ is chosen such that $|\mathcal S_n\rangle$ are among its eigenstates (for example, any local Hamiltonian that is a function solely of $\sigma^z_i$ is a possible $H'$). $H_{\rm e}$ is designed such that the first term annihilates any of the states $|\mathcal S_n\rangle$, so that their energy is set by $H'$.  While a similar scenario arises for the Hamiltonian~\eqref{eq: H} owing to the fact that $H_\lambda|\mathcal S_n\rangle=0$, we stress that this model is \textit{not} of the form \eqref{eq: He}.  In particular, as we show in Appendix~\ref{sec: eigenstate proof}, $|\mathcal S_n\rangle$ are not frustration-free zero-energy states of $H_\lambda$, while they are for $H_{0}$ by design.  However, one can view Eq.~\eqref{eq: He} as defining a class of perturbations to the Hamiltonian \eqref{eq: H} under which the scarred eigenstates $|\mathcal S_n\rangle$ remain stable.

Finally, it is interesting to note that, while projectors $\mathcal P^\prime_i$ analogous to Eq.~\eqref{eq: Pi def} can also be defined for the states $|\mathcal S^\prime_n\rangle$ [Eq.~\eqref{eq: Snp}] by applying the transformation $G$, the projectors $\mathcal P_i$ \textit{do not} annihilate the states $|\mathcal S^\prime_n\rangle$.  Thus, while Hamiltonians of the form \eqref{eq: Hu} and \eqref{eq: He} can be constructed for each tower of states individually, it is apparently not possible to embed \textit{both} towers of states in the ground state manifold of a local Hamiltonian, or among the excited states of another Hamiltonian using the embedding technique of Ref.~\cite{Shiraishi17}.  Despite this, both towers of states appear as eigenstates of the local Hamiltonian \eqref{eq: H}.

\section{Structure of the Scarred Eigenstate Towers}
\label{sec: Structure of the Scarred Eigenstate Towers}

Having defined two sets of scarred eigenstates of Eq.~\eqref{eq: H} and elucidated some of their atypical properties, we now consider their structure in more detail.  We focus as before on the states $|\mathcal S_n\rangle$.  These states form a ``tower of states" in the sense that they are obtained by repeated application of an operator $Q^\dagger$ [Eq.~\eqref{eq: Sn b}] on a ``lowest-weight" state $|\mathcal S_0\rangle = |\Omega\rangle$.  The ``tower" truncates when the ``highest-weight" state $|\mathcal S_{L/2-1}\rangle$ is reached, i.e.~$Q^\dagger |\mathcal S_{L/2-1}\rangle=0$.   Given that $Q^\dagger$ creates a quasiparticle (i.e.~a magnon), one might expect that the operator $Q$ destroys one.  If this were the case, then we could repeatedly apply $Q$ to the ``highest-weight" state $|\mathcal S_{L/2-1}\rangle$ until we reach the lowest-weight state $|\mathcal S_0\rangle$.  This is the case in the spin-1 $XY$ model studied in Ref.~\cite{Schecter19}; in that model, the equivalents of $Q,Q^\dagger$ are generators of an emergent SU(2) algebra, and the scarred states form a particular representation of this algebra.  We now show that this \textit{is not} true in the case of the present model, Eq.~\eqref{eq: H}.  Instead, we will show that a different, \textit{nonlocal} operator $\tilde Q$ can be applied repeatedly to the highest-weight state until the lowest-weight state is reached.  Since the raising and lowering operators for the tower of states are not adjoints of one another, this tower of states does not form a representation of an SU(2) algebra.

Before writing down the nonlocal operator $\tilde Q$, let us consider why a local operator does not allow one to ``descend" the tower.  First, note that the operator $Q$ \textit{does} annihilate a single magnon, i.e.~$Q\, |\mathcal S_1\rangle\propto |\mathcal S_0\rangle$.  This is due to the fact that the lone $1$ present in any configuration entering $|\mathcal S_1\rangle$ is surrounded by $0$s.  However, for OBC one can show that $Q\, |\mathcal S_2\rangle$ is not proportional to $|\mathcal S_1\rangle$.  Instead, one finds
\begin{align}
Q\, |\mathcal S_2\rangle
\propto 
\cdots 
\!+\!
\frac{L-4}{L-5}\, |010\dots0\rangle
\!-\!
|0010\dots 0\rangle
\!+\!
\cdots,
\end{align}
wherein not all states are weighted with equal amplitudes (up to signs): configurations with a magnon near the boundary obtain an enhanced weight due to the freezing of the edge spins under OBC. This unwanted enhancement arises because there are $L-4$ possible locations for the second magnon when the first magnon is at site $2$ or $L-1$, as opposed to $L-5$ when the first magnon is in the bulk of the chain.  A more severe problem arises regardless of boundary conditions for the states $|\mathcal S_n\rangle$ with $n\geq 3$.  For example,
\begin{align}
\label{eq: QS3}
\begin{split}
Q\, |\mathcal S_3\rangle
&\propto
\cdots
+
\frac{L-7}{L-8}\,
|\dots 01010 \dots\rangle\\
&
\qquad\qquad\qquad
-
|\dots 010010 \dots\rangle
+\cdots.
\end{split}
\end{align}
Here an undesired weight enhancement occurs for configurations with two closely packed magnons in the bulk of the chain, where the third magnon can be distributed over $L-7$ (for OBC) rather than $L-8$ sites for nonoverlapping magnons.

The preceding considerations demonstrate that a local operator cannot connect the state $|\mathcal S_n\rangle$ to the state $|\mathcal S_{n-1}\rangle$.  For example, to connect $|\mathcal S_3\rangle$ to $|\mathcal S_2\rangle$, it is not enough to simply annihilate a magnon---one has to keep track of where the other two magnons are, regardless of how far away they are from the third, in order to weight the resulting configurations correctly.  However, there is a nonlocal operator that does the job.  This operator is given by
\begin{align}
\label{eq: qt}
\tilde Q \!=\!
\sum^{i_L}_{i=i_1}\sum^{\frac{L}{2}-1}_{j=0}
(-1)^{i+j}P^0_{i-1}
\!\left[\prod^{i+2j}_{k=i}\sigma^{(-1)^{k-i+1}}_{k}\!\right]\!P^0_{i+2j+1},
\end{align}
where we use the shorthand $\sigma^{(-1)^j}_{k} =\left[\sigma^x_{k}+i(-1)^j\sigma^y_k\right]/2$, which assigns a spin raising or lowering operator to site $k$ depending on the parity of $j$.  Rather than annihilating a single magnon, $\tilde Q$ finds N\'eel domains of odd length with 1s at their left and right ends and, if the domain is flanked by 0s, flips all spins in the domain. This decreases by one the number of 1s in the domain, indicating that $\tilde Q$ decreases the number of magnons by one and the number of domain walls by two, as it should.  This is precisely what is needed to fix the weighting problem. Consider, e.g., Eq.~\eqref{eq: QS3}: while $|\dots 01010 \dots\rangle$ has $L-7$ possible locations for an additional isolated magnon and $|\dots 010010 \dots\rangle$ has $L-8$, one can also add a magnon to the former configuration by flipping the $01010$ domain to $10101$.  However, there are \textit{two} ways to add a magnon to the $010010$ domain in the latter configuration, since any of the two $010$ domains can be flipped to $101$.  Thus, flipping N\'eel domains instead of individual spins enhances the weight of previously underweighted configurations in order to maintain the equal weighting of all configurations in the states $|\mathcal S_n\rangle$.  Consequently, one finds that $\tilde Q\, |\mathcal S_n\rangle \propto |\mathcal S_{n-1}\rangle$ for \textit{any} $n$.  We note in passing that an operator $\tilde{Q}^\prime$ analogous to $\tilde{Q}$ for the states $|\mathcal S^\prime_{n}\rangle$ [Eq.~\eqref{eq: Snp}] can be obtained from Eq.~\eqref{eq: qt} by applying the transformation $G$.

It is intriguing to note that a similar structure might be at work in the tower of scarred eigenstates of the AKLT model derived in Ref.~\cite{Moudgalya18a}, which also feature emergent kinetic constraints like the ones arising in this work.  Ref.~\cite{Moudgalya18a} identified a local ``raising" operator for the tower of states, but the adjoint of this ``raising" operator does not act as a lowering operator.  Based on the similarities with the states studied in this paper, one might suspect the existence of a nonlocal operator like Eq.~\eqref{eq: qt} in the AKLT model; this question is an interesting subject for future work.

\section{Initial States and Finite-Entanglement Revivals}
\label{sec: Initial States and Finite-Entanglement Revivals}
We now turn to the identification of a class of initial states suitable to observe revivals.
A generic initial state that projects the quantum dynamics onto the set of scarred states $|\mathcal S_n\rangle$ is
\begin{align}
\label{eq: generic initial}
|\Psi\rangle=\sum_{n}c_n\, |\mathcal S_n\rangle.
\end{align}
This state has the following three properties, which also turn out to be sufficient for a state to be of the form \eqref{eq: generic initial}: 1) it satisfies the Fibonacci constraint, 2) since each $|\mathcal S_n\rangle$ has magnetization $2n-L$ in the $z$-basis, every configuration in $|\Psi\rangle$ with magnetization $2n-L$ appears with equal weight $|c_n|^2$, and 3) each 1 (i.e., up spin) carries a phase $\pm 1$ depending on whether it sits on an even or odd site.  As with all the results in this paper, an analogous statement holds for the states $|\mathcal S^\prime_n\rangle$, which can be obtained by transforming all states and operators by the transformation $G$. We now discuss one interesting subset of this class of states that is generically entangled, unlike the initial states in Ref.~\cite{Schecter19} and in the original Rydberg experiment~\cite{Bernien17}.  We then discuss an algebraic prescription for constructing such states and explain how it applies to the aforementioned entangled initial states.

\subsection{Rokhsar-Kivelson Initial States}
\label{sec: Rokhsar-Kivelson Initial States}

One class of initial states satisfying all three of the above conditions is
\begin{subequations}
\label{eq: RK}
\begin{align}
|\xi\rangle
=
\frac{1}{\sqrt{Z\left(|\xi|^2\right)}}
\prod^{i_L}_{i=i_1}\left[1+(-1)^i\, \xi\,P^0_{i-1}\, \sigma^+_i\, P^0_{i+1}\right]|\Omega\rangle,
\end{align}
which consists of a superposition of all possible spin flips starting from the ``vacuum" state $|\Omega\rangle$.  Each flipped spin in the superposition carries the necessary staggered phase factor as well as an arbitrary prefactor $\xi\in\mathbb C$, such that any configuration with $n$ flipped spins has weight $|\xi|^{2n}/Z\left(|\xi|^2\right)$. (We will see later that the phase of $\xi$ is physically unimportant, so that we can take $\xi$ to be real and positive.) The normalization factor
\begin{align}
\label{eq: Z}
Z\left(|\xi|^2\right)&=\sum^{L/2-1}_{n=0}|\xi|^{2n}\mathcal N(n)\\
&=\frac{\left(1+\sqrt{1+4 \left| \xi \right| ^2}\right)^{L}\!\!-\!\left(1-\sqrt{1+4 \left| \xi \right| ^2}\right)^{L}}{2^L\sqrt{1+4 \left| \xi \right| ^2}},\nonumber
\end{align}
\end{subequations}
where we have assumed OBC. By construction, the state $|\xi\rangle$ is of the form \eqref{eq: generic initial}, with 
\begin{align}
\label{eq: cn}
c_n = \xi^n\sqrt{\frac{\mathcal N(n)}{Z\left(|\xi|^2\right)}}.
\end{align}
Furthermore, $|\xi\rangle$ is area-law entangled because it can be written as a finite-bond-dimension MPS (see Appendix~\ref{sec: xi MPS}).  We also emphasize that for generic $\xi$, the state $|\xi\rangle$ has finite energy density with respect to $H$---for example, at $L=12$, $\langle \xi=1|H|\xi=1\rangle=-1.28333$, whereas the extremal eigenstates have energies $\sim \pm 15$.

Remarkably, Eq.~\eqref{eq: RK} is equivalent (up to a product of single-site unitary transformations) to the family of Rokhsar-Kivelson (RK) wavefunctions~\cite{Rokhsar88} shown by Lesanovsky~\cite{Lesanovsky11} to be ground states of a family of Hamiltonians relevant to experiments on Rydberg-blockaded atomic ensembles.  The Hamiltonian whose ground state is the state $|\xi\rangle$ is given by
\begin{align}
\label{eq: H xi}
H_{\xi} = \sum^{i_L}_{i=i_1} P^0_{i-1}\left[\xi^{-1}\, P^1_{i}+\xi\, P^0_{i}-(-1)^i\, \sigma^x_i\right]P^0_{i+1},
\end{align}
which can be obtained from Eq.~(4) of Ref.~\cite{Lesanovsky11} by applying the unitary transformation $\prod_{i\text{ even}}\sigma^z_i$.  Thus, to prepare the state $|\xi\rangle$ it suffices to prepare the ground state of $H_{\xi}$, e.g., by quasi-adiabatic ramps, similar to how the N\'eel state was prepared in Ref.~\cite{Bernien17}.

The state $|\xi\rangle$ can be viewed as containing a distribution of $\pi$-momentum magnons parameterized by $|\xi|$.  In particular, the normalization factor $Z\left(|\xi|^2\right)$, Eq.~\eqref{eq: Z}, can be interpreted as a grand canonical partition function for a gas of $\pi$-magnons at infinite temperature and finite fugacity $|\xi|^2$.  It immediately follows that the expansion coefficients $|c_n|^2$, Eq.~\eqref{eq: cn}, can be interpreted as the classical grand-canonical probability of the system occupying a state with $n$ $\pi$-magnons.  Thus, by tuning $|\xi|$, one can effectively tune the initial distribution of $\pi$-magnons and subsequently follow the quantum evolution of this distribution.

\begin{figure}[t!]
\begin{center}
\includegraphics[width=.85\columnwidth]{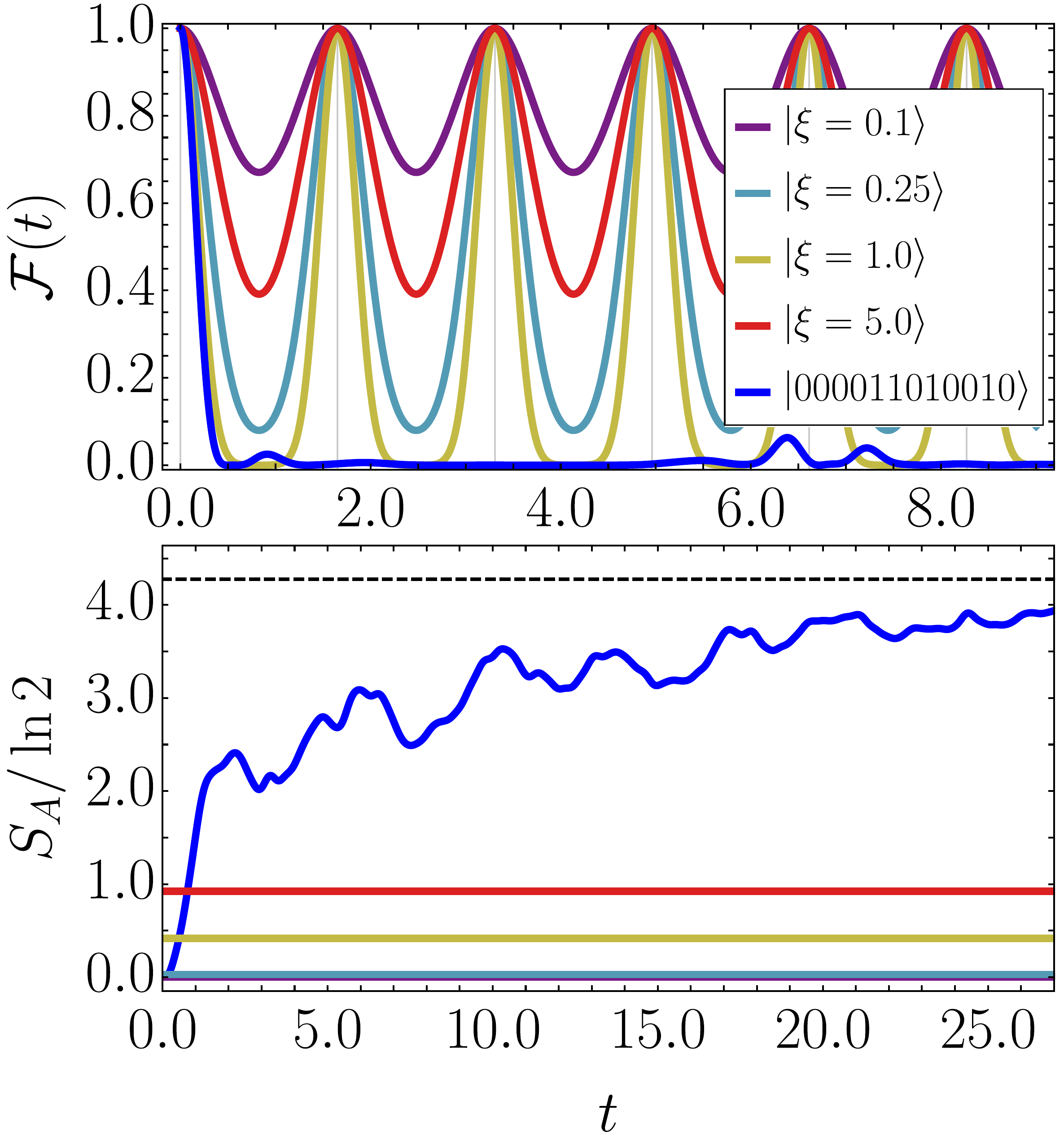}\\
\caption{
Dynamics of the many-body fidelity $\mathcal F(t)=|\langle\Psi(t)|\Psi(0)\rangle|^2$ (top) and the half-chain entanglement entropy $S_A$ (bottom) for various initial states at $L=12$. (Other parameters are the same as Figs.~\ref{fig: lvl} and \ref{fig: ee}.)  For the RK states $|\xi\rangle$, the fidelity dynamics [for which Eq.~\eqref{eq: F} is plotted] exhibits exact periodic revivals with period $2\pi/\Omega$, while the entanglement remains constant in time.  The finite value of the fidelity between revivals for some values of $\xi$ is a finite-size effect.   For an arbitrary product state in the $z$-basis with a comparable energy density, the fidelity rapidly decays to zero while the entanglement grows close to the random-state value, Eq.~\eqref{eq: S_A ran} (black dashed line).
}
\label{fig: dynamics}
\end{center}
\end{figure}

Since $|\xi\rangle$ is of the form \eqref{eq: generic initial}, its evolution is simply
\begin{align}
\label{eq: evol}
|\xi(t)\rangle
&=
\sum_{n}c_n\, e^{-i E_n t}|\mathcal S_n\rangle \nonumber \\
&\propto
\frac{1}{\sqrt{Z\left(|\xi|^2\right)}}\sum_n \sqrt{\mathcal N(n)}\, (e^{-i \Omega t}\xi)^n|\mathcal S_n\rangle\\
&=|e^{-i\Omega t} \xi\rangle, \nonumber
\end{align}
up to an overall phase, with $c_n$ and $E_n$ defined in Eqs.~\eqref{eq: cn} and \eqref{eq: En}, respectively, and where $\Omega \equiv 2\Delta-4J$.
From this expression, we immediately see that the state $|\xi\rangle$ returns to itself with period $2\pi/\Omega$ under time evolution with $H$.  This periodic behavior is reflected in the many-body fidelity under evolution with $H$,
\begin{align}
\label{eq: F}
\mathcal F_\xi(t) = |\langle \xi(t)|\xi\rangle|^2 = \left|\frac{Z\left(e^{i\Omega t}|\xi|^2\right)}{Z\left(|\xi|^2\right)}\right|^2.
\end{align}
In the thermodynamic limit and for any finite $\xi$, the above expression approaches a function that equals $1$ at integer multiples of $2\pi/\Omega$ and $0$ everywhere else. These exact periodic revivals constitute the final definitive hallmark of QMBS.

We plot the fidelity and entanglement dynamics for various initial states in Fig.~\ref{fig: dynamics}.  As expected, the states $|\xi\rangle$ display exact revivals with period $2\pi/\Omega$ (in agreement with the analytical expression \eqref{eq: F} to numerical precision), while a generic product state in the $z$-basis does not. Moreover, while the half-chain entanglement entropy for an initial product state grows rapidly with time and approaches the value $S^\mathrm{ran}_A$, Eq.~\eqref{eq: S_A ran}, it remains constant for the initial states $|\xi\rangle$.  This unusual feature highlights the fact that the evolution of the states $|\xi\rangle$ under $H$ is exceptionally simple despite their finite entanglement; as evident in Eq.~\eqref{eq: evol}, time evolution merely rotates the phase of $\xi$ at a frequency $\Omega$.

\begin{figure}[t!]
\begin{center}
\includegraphics[width=.87\columnwidth]{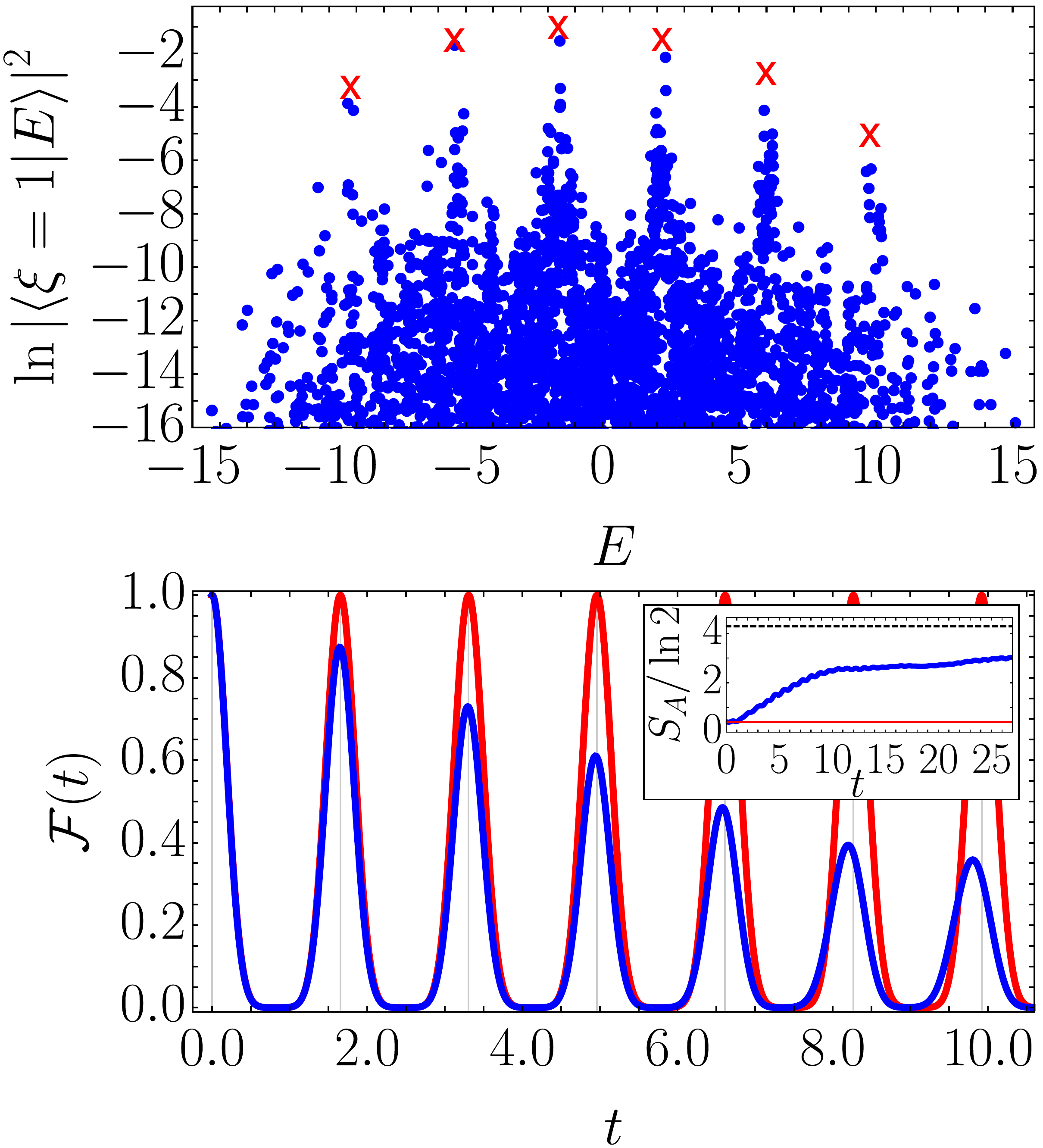}\\
\caption{
Persistence of scarred dynamics upon perturbing Eq.~\eqref{eq: H} with $h\sum_i\sigma^x_i$ for $h=0.15$ (all other parameters are as in Fig.~\ref{fig: dynamics}).  Top: Overlap of the initial state $|\xi=1\rangle$ with each exact eigenstate of the perturbed model; red crosses indicate the energies and overlaps of the scarred eigenstates in the unperturbed model.  Bottom: Dynamics of the fidelity and entanglement entropy (inset) for the perturbed (blue) and unperturbed (red) models with the initial state $|\xi=1\rangle$.  All quantities shown bear a striking resemblance to their counterparts in the PXP model~\cite{Turner17}.
}
\label{fig: pert}
\end{center}
\end{figure}

It is interesting to consider the effects of small perturbations to the model \eqref{eq: H} on the periodic dynamics discussed here.  Adding a generic perturbation, e.g.~$h\sum_i\sigma^x_i$, which breaks the U(1) symmetry of Eq.~\eqref{eq: H}, removes the exact scarred eigenstates constructed in this paper.  Nevertheless, for sufficiently small $h$, the initial state $|\xi\rangle$ still primarily overlaps with exact eigenstates that are concentrated in a finite window around the ``scar" energies $E_n$, similar to what has been observed for the N\'eel state in the PXP model~\cite{Turner17} (see Fig.~\ref{fig: pert}, top panel).  This enables coherent fidelity dynamics with \textit{imperfect} revivals over a finite lifetime, also similar to what is seen in the PXP model (see Fig.~\ref{fig: pert}, bottom panel).  A general argument based on Lieb-Robinson bounds that is consistent with these results has been made in Ref.~\cite{Lin19}; an interesting subject for future work would be to extend the results of that work by determining tighter bounds on the lifetime of the imperfect revivals for perturbations of specific models with exact scars like the one considered in this paper.

\subsection{Initial States from Projected SU(2) Rotations}
We now discuss a related strategy for obtaining initial states satisfying (1)--(3).  This strategy hinges on the use of rotations generated by
\begin{align}
\label{eq: su(2)}
\mathcal J^{x/y}=\frac{1}{2}\sum^{i_L}_{i=i_1}(-1)^i\, \sigma^{x/y}_i,\indent \mathcal J^z=\frac{1}{2}\sum^{i_L}_{i=i_1}\sigma^z_i,
\end{align}
which can readily be shown to satisfy the SU(2) algebra $[\mathcal J^\alpha,\mathcal J^\beta]=i\, \epsilon^{\alpha\beta\gamma}\,\mathcal J^\gamma$.  Using these generators, we can define an auxiliary tower of states
\begin{align}
\label{eq: aux}
|S_n\rangle\propto(\mathcal J^+)^n\, |\Omega\rangle,
\end{align}
satisfying $|\mathcal S_n\rangle = \mathcal P_{\rm fib}|S_n\rangle$ for any $n\leq L/2$, where $\mathcal J^\pm=\mathcal J^x\pm i\,\mathcal J^y$ and $\mathcal P_{\rm fib}$ is the projector onto states obeying the Fibonacci constraint.  While they are not eigenstates of the Hamiltonian \eqref{eq: H}, the auxiliary states \eqref{eq: aux} \textit{do} form a representation of SU(2), as can be verified by computing the action of $\bm{\mathcal J}\cdot\bm{\mathcal J}=\frac{1}{2}(\mathcal J^+\mathcal J^-+\text{H.c.})+(\mathcal J^z)^2$ on any normalized $|S_n\rangle$. We find
\begin{align}
\bm{\mathcal J}\cdot\bm{\mathcal J}\, |S_n\rangle = l(l+1)\, |S_n\rangle,\ l=\begin{cases}\frac{L}{2}-1 & \text{OBC}\\ \frac{L}{2} & \text{PBC}\end{cases},
\end{align}
where $2l+1$ is the number of states in the auxiliary tower \eqref{eq: aux}, which depends on boundary conditions due to the definitions \eqref{eq: su(2)}.

We emphasized above that the states \eqref{eq: aux} are not eigenstates of Eq.~\eqref{eq: H}, and in Sec.~\ref{sec: Structure of the Scarred Eigenstate Towers} that the scarred eigenstates $|\mathcal S_n\rangle$ cannot form a representation of any SU(2) algebra; nevertheless, the SU(2) algebra \eqref{eq: su(2)} provides a prescription for generating initial states of the form \eqref{eq: generic initial}, as we now explain.  Any state in the representation of the algebra \eqref{eq: su(2)} spanned by the auxiliary states $|S_n\rangle$ satisfies conditions (2) and (3) by construction: configurations with fixed magnetization have equal weight, and each up spin carries a staggered sign. The remaining condition, (1), is satisfied once the projector $\mathcal P_{\rm fib}$ is applied.  We conclude that any state of the form
\begin{align}\label{eq: generic initial P}
|\Psi\rangle \propto \mathcal P_{\rm fib}\sum_{n}\tilde c_n |S_n\rangle,
\end{align}
once normalized, is automatically of the form \eqref{eq: generic initial}.  Such states can generically be obtained by acting with an SU(2) rotation on any state belonging to the appropriate SU(2) representation and subsequently applying $\mathcal P_{\rm fib}$.

As an application of this concept, we demonstrate that the RK states \eqref{eq: RK} are of the form \eqref{eq: generic initial P}.  First, we rewrite the RK state as
\begin{align}
|\xi\rangle
=
\frac{1}{\sqrt{Z\left(|\xi|^2\right)}}\, \mathcal P_{\rm fib}
\prod^{i_L}_{i=i_1}\left[1+(-1)^i\, \xi\, \sigma^+_i\, \right]|\Omega\rangle.
\end{align}
We can then show that
\begin{align}
\label{eq: RK-R}
\prod^{i_L}_{i=i_1}\left[1+(-1)^i\, \xi\, \sigma^+_i\, \right]|\Omega\rangle \propto \mathcal R(\phi,\theta,\psi)\, |\Omega\rangle,
\end{align}
where $\mathcal R(\phi,\theta,\psi)$ is an SU(2) rotation parameterized by Euler angles $\phi,\theta,\psi$ as follows:
\begin{align}
\mathcal R(\phi,\theta,\psi)=e^{i\, \phi\, \mathcal J^z}e^{i\, \theta\, \mathcal J^x}e^{i\, \psi\, \mathcal J^z}.
\end{align}
A straightforward calculation finds that 
\begin{align}
\xi = \tan(\theta/2)\, e^{i(\phi+\pi/2)}
\end{align}
in Eq.~\eqref{eq: RK-R}.  Note that the Euler angle $\psi$ does not enter as it merely results in an overall phase factor.  Since $|\Omega\rangle=|S_0\rangle$ and the states $|S_n\rangle$ form a basis for a representation of the SU(2) algebra \eqref{eq: su(2)}, Eq.~\eqref{eq: RK-R} immediately implies the desired result.

As a final comment, we stress that the projected SU(2) structure of the states~\eqref{eq: generic initial P} \textit{does not} imply an interpretation of the periodic dynamics in terms of the precession of an SU(2) spin, even at the level of the auxiliary tower of states \eqref{eq: aux}.  If such a picture did hold, it would be possible to rotate the precession axis [which for us is the $z$-axis due to the structure of the Hamiltonian \eqref{eq: H}] by adding to $H$ some linear combination of the generators \eqref{eq: su(2)}.  This procedure is discussed in Ref.~\cite{Schecter19}, where it is made possible by the fact that the scarred \textit{eigenstates} constructed there form a representation of SU(2). Such a rotation of the precession axis is precluded in the present case by the fact that the scarred eigenstates constructed here do not form an SU(2) representation, as discussed in Sec.~\ref{sec: Structure of the Scarred Eigenstate Towers}.  Instead, a special quantization axis is selected by the Fibonacci constraint: $[\mathcal J_z,\mathcal P_{\rm fib}]=0$, which singles out the $z$-axis as the preferred axis of rotation.

\section{Conclusions}
\label{sec: Conclusions}

In this paper we have demonstrated the existence of exact quantum many-body scars in the nonintegrable model $H$ defined in \eqref{eq: H}.  The scarred many-body eigenstates in this model differ in several important respects from previous examples of QMBS.  First, the model possesses two towers of scarred eigenstates, related by a $\mathbb Z_2$ transformation that does not commute with $H$.  Second, the scarred towers of states possess an unusual structure where there is an asymmetry between creating and destroying a magnon---the former is accomplished with a local operator, whereas the latter is accomplished with a nonlocal operator.  Third, we find that the scarred eigenstates give rise to periodic revivals when the system is prepared in a family of \textit{entangled} initial states.  This scenario differs from most studies of QMBS to date, as typically the initial state is taken to be a product state.  However, we believe that a state of the form \eqref{eq: RK} will generically allow for the observation of revivals in systems with QMBS for which a quasiparticle picture holds either exactly or approximately. For example, for the spin-1 $XY$ model studied in Ref.~\cite{Schecter19}, one can replace the $P^0\sigma^+P^0$ term in Eq.~\eqref{eq: RK} with $(S^+_i)^2$ and set $\xi=1$ to obtain the product state used to elicit revivals in that model. 

We have also investigated in Sec.~\ref{sec: Rokhsar-Kivelson Initial States} the effect of perturbing the model \eqref{eq: H} by breaking the conservation of $n_{\rm DW}$.  For a sufficiently small perturbation, we find that the perviously exact periodic revivals of the chosen initial states become highly coherent \textit{approximate} revivals with a finite lifetime, similar to what is seen in the PXP model.  A more detailed study of such perturbations and their induced timescales will be an interesting subject for future studies.

The scarred eigenstates in this model share several features with the scarred eigenstates of the AKLT chain that were found in Ref.~\cite{Moudgalya18a}.  For example, the scarred eigenstates of the AKLT chain also have emergent kinetic constraints that prevent quasiparticles from being created on neighboring sites.  This also gives rise to an asymmetry in the AKLT tower of states similar to the one studied here.  Another interesting problem for future work is therefore to use the insights gained in the present work to better understand QMBS in the AKLT model.

It will also be interesting to consider whether other states of the form \eqref{eq: generic initial} can be prepared by local means, i.e.~to understand how much control can be exerted over the amplitudes $c_n$.  Tailoring these amplitudes is equivalent to designing a grand-canonical distribution function for the $\pi$-magnons, which in turn strongly influences the ensuing periodic dynamics.  Mapping out the space of possible $c_n$ is equivalent to mapping out the full set of locally preparable initial states that give rise to scarred many-body dynamics.

Finally, we note that the model \eqref{eq: H} maps exactly onto a $\mathbb Z_2$ lattice gauge theory coupled to fermionic matter, see Ref.~\cite{Borla19}.  The present work thus provides an opportunity to strengthen the connection between QMBS and lattice gauge theories suggested in Refs.~\cite{Surace19,Magnifico19}, which presents an interesting pathway for future progress.  Furthermore, schemes for realizing $\mathbb Z_2$ gauge theories coupled to fermionic matter with cold atoms in optical lattices have been proposed in, e.g., Ref.~\cite{Barbiero19}, and steps have recently been taken towards their experimental implementation, see Refs.~\cite{Gorg19,Schweizer19}. These ongoing developments suggest the possibility of new experimental platforms for the observation and study of QMBS.

Moreover, it was recently shown~\cite{Ostmann19} that the model \eqref{eq: H} can be realized directly in Rydberg atomic lattices in the ``antiblockade" regime. This provides an appealing path to experimental realization of the quench protocol discussed in this paper. Indeed, the Hamiltonian \eqref{eq: H xi} can be realized approximately in Rydberg atomic lattices in the blockade regime, as discussed in Ref.~\cite{Lesanovsky11}. Thus, the quench protocol suggested in our work can be (approximately) achieved by (approximately) preparing the ground state of Eq.~\eqref{eq: H xi}, e.g.~by quasiadiabatic ramps into the strong nearest-neighbor blockade regime, and subsequently quenching into the antiblockade regime where Eq.~\eqref{eq: H} is realized.  The ensuing dynamics should yield approximate revivals similar to those in Fig.~\ref{fig: pert}.

\textit{Note added}---In the final stages of preparing this manuscript there appeared Ref.~\cite{Chattopadhyay19}, which discusses finite-entanglement revivals in a variant of the model studied in Ref.~\cite{Schecter19} and obtains similar results to the ones reported here.

\acknowledgments{
We thank Shenglong~Xu for useful discussions on matrix product operators and
Sanjay~Moudgalya for bringing to our attention the connection between our initial states and the class of ground states studied in Ref.~\cite{Lesanovsky11}.
We also thank Fabian Grusdt for pointing out the connection to lattice gauge theories, and Juan P.~Garrahan for pointing out the
possible realization of the model \eqref{eq: H} in antiblockaded Rydberg atomic lattices.
This work is supported by Microsoft and the Laboratory for Physical Sciences. 
T.I. acknowledges a JQI Postdoctoral Fellowship and Iowa State University startup funds.
}

\begin{widetext}

\appendix

\section{Proof that the states \eqref{eq: Sn} are eigenstates}
\label{sec: eigenstate proof}

The proof that the states $|\mathcal S_n\rangle$ are eigenstates of $H$ follows closely the analogous one presented in Ref.~\cite{Moudgalya18a}.  Since the states $|\mathcal S_n\rangle$ are automatically eigenstates of $H_z$ and $H_{zz}$ [see Eqs.~\eqref{eq: H} and \eqref{eq: En}], it suffices to show that $H_\lambda |\mathcal S_n\rangle=0$.   We begin by considering the action of $H_\lambda$ on $|\mathcal S_1\rangle$ and $|\mathcal S_2\rangle$, before considering states with more magnons.  For simplicity we focus on the case of PBC, although at the end we comment on how the proof is modified for OBC.  

Adopting notation analogous to that of Ref.~\cite{Moudgalya18a}, we write
\begin{subequations}
\begin{align}
|\mathcal S_1\rangle &\propto \sum^{L}_{j=1}(-1)^{j}\, |M_{j}\rangle, \label{eq: S1} \\ 
|\mathcal S_2\rangle &\propto \sum^{L}_{\substack{j_1,j_2=1\\ |j_1-j_2|>1}}(-1)^{j_1+j_2}\, |M_{j_1} M_{j_2}\rangle =\sum^{L}_{j=1}\sum^{L-2}_{m=2}(-1)^m\, |M_j M_{j+m}\rangle, \label{eq: S2}
\end{align}
where $|M_j\rangle$ is a state containing one magnon on site $j$, so that
\begin{align}
\label{eq: Mj}
\begin{split}
|M_{j}\rangle &\equiv |0\dots0\underset{j}{1}0\dots0\rangle\\
|M_{j_1}M_{j_2}\rangle &\equiv |0\dots0\underset{j_1}{1}0\dots0\underset{j_2}{1}0\dots0\rangle
\end{split}
\end{align}
\end{subequations}
and so on for states containing more magnons.  Next we compute
\begin{align}
\label{eq: HM}
H_\lambda |M_j\rangle = 2\left(|0\dots1\underset{j}{1}0\dots0\rangle +|0\dots\underset{j}{1}10\dots0\rangle\right)\equiv 2\left(|N_{j-1}\rangle+|N_j\rangle\right).
\end{align}
We will call $|N_{j-1}\rangle$ and $|N_{j}\rangle$ the ``backward" and ``forward" scattering states, respectively.
Applying $\sum^{L}_{j=1}(-1)^j$ to both sides of the above delivers $H_\lambda|\mathcal S_1\rangle=0$ due to a cancellation of scattering states $|N_{j}\rangle$ between $|M_j\rangle$ and $|M_{j+1}\rangle$.
We can now apply the scattering rule \eqref{eq: HM} to $|\mathcal S_2\rangle$.  We compute
\begin{align}
H_\lambda |M_j M_{j+m}\rangle
=
2\times
\begin{cases}
|N_{j-1}M_{j+m}\rangle+|N_{j}M_{j+m}\rangle+|M_jN_{j+m-1}\rangle+|M_jN_{j+m}\rangle & 3 \leq m \leq L-3\\
|N_{j-1}M_{j+2}\rangle+|M_jN_{j+2}\rangle & m=2 \\
|N_{j}M_{j-2}\rangle+|M_jN_{j-3}\rangle & m=L-2
\end{cases}
,
\end{align}
where configurations with $m=2,L-2$ are distinguished from the others because they contain the motif $\dots 01010 \dots$, where scattering into the central ``0" site is forbidden by the form of $H_\lambda$ [see discussion around Eq.~\eqref{eq: scattering}].  For the $m=L-2$ case above we used the fact that $j+L-2\equiv j-2$ mod L.  We then obtain
\begin{subequations}
\begin{align}
\begin{split}
\label{eq: HS2a}
H_{\lambda}|\mathcal S_2\rangle
&\propto \sum^{L}_{j=1}\sum^{L-3}_{m=3}(-1)^m\Big( |N_{j}M_{j+m}\rangle+|M_jN_{j+m-1}\rangle+|N_{j-1}M_{j+m}\rangle+|M_jN_{j+m}\rangle \Big)\\
&\qquad +\sum^{L}_{j=1} \Big(|N_{j-1}M_{j+2}\rangle+|M_jN_{j+2}\rangle+|N_{j}M_{j-2}\rangle+|M_jN_{j-3}\rangle\Big)
\end{split}
\\
\label{eq: HS2b}
&=\sum^L_{j=1}\left[\sum^{L-4}_{m=3} (-1)^m \Big(|N_{j-1}M_{j+m}\rangle +|M_{j}N_{j+m}\rangle\Big)+\sum^{L-3}_{m=4}(-1)^m\Big( |N_jM_{j+m}\rangle+|M_{j}N_{j+m-1}\rangle \Big)\right],
\end{align}
\end{subequations}
where in going from \eqref{eq: HS2a} to \eqref{eq: HS2b} we have used the terms on the second line of \eqref{eq: HS2a} to cancel some of the terms in the first line.  We next show that the remaining terms in Eq.~\eqref{eq: HS2b} cancel.  First, we note that
\begin{align}
\sum^L_{j=1}\sum^{L-4}_{m=3} (-1)^m |N_{j-1}M_{j+m}\rangle=\sum^L_{j=1}\sum^{L-4}_{m=3} (-1)^m |N_{j}M_{j+m+1}\rangle=-\sum^L_{j=1}\sum^{L-3}_{m=4} (-1)^m |N_{j}M_{j+m}\rangle,
\end{align}
where in the first equality we redefined $j-1\to j$ and in the second equality we redefined $m+1\to m$.  Second, we note that
\begin{align}
\sum^{L-3}_{m=4}(-1)^m |M_{j}N_{j+m-1}\rangle=-\sum^{L-4}_{m=3}(-1)^m |M_{j}N_{j+m}\rangle
\end{align}
upon redefining $m-1\to m$.  We conclude that Eq.~\eqref{eq: HS2b} is identically zero, so that $H_\lambda|\mathcal S_2\rangle=0$ as desired.  Note that in order to prove this we needed to shift the summation indices---in other words, the cancellation of scattering terms does not occur locally, but rather only occurs once all sites have been summed over.  This indicates that $|\mathcal S_2\rangle$ is \textit{not} a frustration-free eigenstate of $H_\lambda$.  The same holds true for the remaining states $|\mathcal S_n\rangle$.

Next we consider the action of $H_\lambda$ on
\begin{align}
\label{eq: SnM}
|\mathcal S_n\rangle \propto \sum_{\substack{\{j_p\}^n_{p=1}\\ |j_p-j_{p+1}|>1}}\!\!\!\!(-1)^{\sum^n_{p=1}j_p}\, |M_{j_1}\dots M_{j_n}\rangle.
\end{align}
Applying the scattering rule \eqref{eq: HM}, we see that
\begin{align}
H_\lambda |M_{j_1}\dots M_{j_k}\dots M_{j_n}\rangle \propto \sum_{\ell_k}|M_{j_1}\dots N_{\ell_k}\dots M_{j_n}\rangle,
\end{align}
where $\ell_k=j_k$ and/or $j_k-1$ depending on the coordinates of the surrounding magnons.  The scattering states for $M_{j_k}$ are then canceled by other terms in the sum in Eq.~\eqref{eq: SnM} where $M_{j_k}$ is replaced by $M_{j_k\pm 1}$, which necessarily come with a relative minus sign.  It is important to note that $L$ must be even in order for these cancellations to occur when $j_k=1$ or $L$.  This is consistent with the fact that $|\mathcal S_n
\rangle$ is an eigenstate of the translation operator with eigenvalue $(-1)^n$, whereas momenta are quantized in half-integer units when $L$ is odd: thus $|\mathcal S_n\rangle$ cannot be an eigenstate of $H$ when $L$ is odd.

Finally, we comment on the case of OBC.  The above argumentation is essentially unchanged, except for the fact that the states $|\mathcal S_n\rangle$ must be defined such that magnons cannot occupy sites $1,L$.  Furthermore, the sum $\sum^{L-2}_{m=2}$ in \eqref{eq: S2} should be replaced with $\sum^{-2}_{m=-j+2}+\sum^{L-j-1}_{m=2}$.  The analysis of the $n$-magnon case still holds, except that now the restriction that $L$ must be even is lifted since the scattering of $M_2$ and $M_{L-2}$ does not interfere.

\section{Derivation of the entanglement spectrum}
\label{sec: es}

To derive the entanglement spectrum in Eq.~\eqref{eq: es} from Eq.~\eqref{eq: Sn cut}, we first apply a unitary transformation to the state $|\mathcal S_n\rangle$ that removes the staggered phase
factor $\sigma(i_1,\dots,i_n)$.  The transformation that does the job is
\begin{align}
U\equiv \prod_{i\ \text{odd}} \sigma^z_i,
\end{align}
which obeys
\begin{align}
U\, \sigma^\pm_i\, U = (-1)^i\, \sigma^\pm_i.
\end{align}
Note that $U$ is a product of single-site rotations and hence will not alter the entanglement spectrum.  From there, we write
\begin{align}
\label{eq: USN 1}
U|\mathcal S_n\rangle 
&= \frac{1}{\sqrt{\mathcal N(L,n)}}
\sum_{\substack{i_1<\dots<i_n\\ |i_p-i_{p+1}|>1\\ i_p\neq 1,L}}
\!\!\!
|\{i_p\}\rangle
\equiv
\!\!\!\!\!
\sum_{\substack{i_1<\dots<i_n\\ |i_p-i_{p+1}|>1\\ i_p\neq 1,L}}
\tilde M_{\{i_p\}_A,\{i_p\}_B}\,
|\{i_p\}_A\rangle\otimes|\{i_p\}_B\rangle, 
\end{align}
where the matrix $\tilde M$ obtained from the matrix $M$ in Eq.~\eqref{eq: Sn cut} by taking the absolute value of each matrix entry.
The matrix $\tilde M$ has dimension $\mathcal D_A\times\mathcal D_B$, where $\mathcal D_{A(B)}$ is the Hilbert space dimension of region $A$ ($B$).
Next, we break the expression \eqref{eq: USN 1} up into pieces as follows:
\begin{align}
\label{eq: USN 2}
U|\mathcal S_n\rangle 
&=
\!\!\!\!\!\!\!\!
\sum_{\substack{i_1<\dots<i_n\\ |i_p-i_{p+1}|>1\\ i_p\neq 1,L}}
\!\!
\sum^K_{k=0}
\tilde M^k_{\{i_p\}^k_A,\{i_p\}^{n-k}_B}\,
|\{i_p\}^k_A\rangle
\!\otimes\!
|\{i_p\}^{n-k}_B\rangle, 
\end{align}
where $\{i_p\}^k_A$ denotes a configuration in subregion $A$ containing $k$ magnons and $\{i_p\}^{n-k}_B$ denotes a configuration in subregion $B$ containing $n-k$ magnons. The upper limit on the second summation is $K=\text{min}(n,\lfloor L_A/2 \rfloor)$, as it is not possible to fit more than $\lfloor L_A/2 \rfloor$ magnons into the region $A$ without violating the kinetic constraint.

We wish to compute the eigenvalues of the $\mathcal D_A\times\mathcal D_A$ matrix $\tilde{\mathcal M}=\tilde M \tilde M^\dagger$ (note that this matrix has the same spectrum as that of $\mathcal M$).  Since every configuration in $U|\mathcal S_n\rangle$ appears with the same amplitude, the matrix elements of $\tilde{\mathcal M}$ for two configurations $a,b$ in region $A$ are given by
\begin{align}
\label{eq: cal M def}
\tilde{\mathcal M}_{ab}=\frac{N^B_{ab}}{\mathcal N(L,n)},
\end{align}
where $N^B_{ab}$ is the number of configurations in region $B$ that appear with both $a$ and $b$ in Eq.~\eqref{eq: USN 1}.
To compute the matrix $\tilde{\mathcal M}$, we observe that $\tilde{\mathcal{M}}=\bigoplus^K_{k=0}\tilde{\mathcal{M}}_k$, where $\tilde{\mathcal{M}}_k=\tilde M^k (\tilde M^k)^\dagger$ (note that $k$ here is an upper index, not a power).  This block-diagonal structure arises from the fact that a configuration in $A$ containing $k$ magnons \textit{must} be paired with a configuration in $B$ containing $n-k$ magnons.  Using Eq.~\eqref{eq: cal M def}, we find that
\begin{align}
\tilde{\mathcal M}_{k}
=
\frac{1}{\mathcal N}
\begin{pmatrix}
m_{1,k}\, \bm{1}_{D_{1,k}\times D_{1,k}} & m_{2,k}\, \bm{1}_{D_{1,k}\times D_{2,k}} \\
m_{2,k}\, \bm{1}_{D_{2,k}\times D_{1,k}} & m_{2,k}\, \bm{1}_{D_{2,k}\times D_{2,k}}
\end{pmatrix},
\end{align}
where $\bm{1}_{A,B}$ is the $A\times B$ matrix with all entries equal to $1$, and where $m_{1,k}$, $m_{2,k}$, $D_{1,k}$, and $D_{2,k}$ are defined in Eqs.~\eqref{d1k}--\eqref{m2k}.  Here, $D_{1,k}$ and $D_{2,k}$ are defined such that $D_{1,k}+D_{2,k}$ is equal to the number of configurations in region $A$ containing exactly $k$ magnons. (Note that $\tilde{\mathcal M}_k$ is a square matrix of dimension $D_{1,k}+D_{2,k}$.)  $D_{1,k}$ is the number of configurations with a $0$ next to the entanglement cut, and $D_{2,k}$ is the number of configurations with a $1$ next to the entanglement cut.  The numbers $m_{1,k}$ and $m_{2,k}$ denote the number of configurations in region $B$ that are allowed depending on whether the last site in $A$ is a $0$ or a $1$, respectively.  That two different numbers are required for the two cases is a consequence of the kinetic constraint prohibiting two magnons from occupying nearest neighbor sites.

The eigenvectors of the matrix $\tilde{\mathcal M}_{k}$ can be taken to be of the form $(c_1\dots c_1\ |\ c_2\dots c_2)^\mathsf{T}$, where $c_1$ is repeated $D_{1,k}$ times and $c_2$ is repeated $D_{2,k}$ times.  Thus, the characteristic polynomial of the matrix $\tilde{\mathcal M}_{k}$ is the same as that of
\begin{align}
\frac{1}{\mathcal N}
\begin{pmatrix}
m_{1,k}\, D_{1,k} & m_{2,k}\, D_{2,k}\\
m_{2,k}\, D_{1,k} & m_{2,k}\, D_{2,k}
\end{pmatrix}.
\end{align}
The eigenvalues of the above matrix are precisely $\lambda_{k,\pm}$, Eq.~\eqref{eq: lambda}.

\section{Derivation of the ``uncle Hamiltonian" \eqref{eq: Hu}}
\label{sec: uncle}

In this Appendix we outline the derivation of the ``uncle Hamiltonian" for the states $|\mathcal S_n\rangle$, which is based on matrix product state techniques.  We refer the reader to, e.g., Ref.~\cite{Schollwock11} for relevant background material.  Our strategy in the derivation is to express the states $|\mathcal S_n\rangle$ as MPSs, and then to examine their few-site reduced density matrices to find projectors whose common null space is spanned by $\{|\mathcal S_n\rangle\}$.  For concreteness we work with OBCs, although boundary conditions are unimportant for the following analysis.  To find the MPS expressions for $|\mathcal S_n\rangle$, we rewrite Eq.~\eqref{eq: Sn} as
\begin{subequations}
\begin{align}
|\mathcal S_n\rangle=\mathcal P_{\rm fib}\bigg(\underbrace{\sum^{L-1}_{i=2}(-1)^i\, \sigma^+_i}_{\equiv\sigma^+_\pi}\bigg)^n|\Omega\rangle,
\end{align}
where
\begin{align}
\mathcal P_{\rm fib} =\left(\frac{1-\sigma^z_1}{2}\right)\prod^{L-1}_{i=2}\left[1-\left(\frac{1+\sigma^z_i}{2}\right)\left(\frac{1+\sigma^z_{i+1}}{2}\right)\right]\left(\frac{1-\sigma^z_L}{2}\right)
\end{align}
\end{subequations}
is the projector onto the sector of the full Hilbert space in which both edge spins are down and the bulk of the chain obeys the Fibonacci constraint.  We then write the state $|\Omega\rangle$ as a bond-dimension-$1$ MPS,
\begin{subequations}
\begin{align}
|\Omega\rangle = \sum_{\{\sigma_i\}^{L}_{i=1}}a_1[\sigma_1]A_2[\sigma_2]\dots A_{L-1}[\sigma_{L-1}]a_L[\sigma_{L}]\ |\{\sigma_i\}^{L}_{i=1}\rangle
\end{align}
where $\sigma_i=0,1$ denotes the physical spin state on site $i$ and
\begin{align}
a_{1,L}[\sigma]=A_i[\sigma]=\delta_{\sigma,0}.
\end{align}
\end{subequations}
Next, we represent the operator $\left(\sigma^+_\pi\right)^n$ as a matrix product operator (MPO) of bond dimension $\chi_n=n+1$, i.e.
\begin{subequations}
\begin{align}
(\sigma^+_\pi)^n = (m_{(\sigma^+_\pi)^n})_1(M_{(\sigma^+_\pi)^n})_2\cdots(M_{(\sigma^+_\pi)^n})_{L-1}(m_{(\sigma^+_\pi)^n})_L,
\end{align}
where the $\chi_n$-dimensional operator-valued boundary vectors
\begin{align}
(m_{(\sigma^+_\pi)^n})_1[\sigma_1,\sigma^\prime_1]_\alpha&=
\delta_{\alpha,1}
[\mathbbm 1]_{\sigma_1,\sigma^\prime_1}
\\
(m_{(\sigma^+_\pi)^n})_L[\sigma_L,\sigma^\prime_L]_\alpha&=
\delta_{\alpha,\chi_n}
[\mathbbm 1]_{\sigma_L,\sigma^\prime_L}
\end{align}
and the operator-valued $\chi_n\times\chi_n$ matrices
\begin{align}
(M_{(\sigma^+_\pi)^n})_i[\sigma_i,\sigma^\prime_i]_{\alpha,\beta}
=
\delta_{\alpha,\beta}\, (-1)^\alpha\, [\mathbbm 1]_{\sigma_i,\sigma^\prime_i}+\delta_{\beta,\alpha+1}\, (-1)^\alpha\, [\sigma^+]_{\sigma_i,\sigma^\prime_i},
\end{align}
\end{subequations}
where $\alpha,\beta=1,\dots,\chi_n$ and $[\mathcal O]_{\sigma_i,\sigma^\prime_i}$ denotes the matrix element $\langle\sigma^\prime_i|\mathcal O|\sigma_i\rangle$ for any single-site operator $\mathcal O$. We next write the Fibonacci projector as a bond-dimension-2 MPO,
\begin{subequations}
\label{eq: Pfib MPO}
\begin{align}
\mathcal P_{\rm fib} = (m_{\mathcal P_{\rm fib}})_1(M_{\mathcal P_{\rm fib}})_2\cdots(M_{\mathcal P_{\rm fib}})_{L-1}(m_{\mathcal P_{\rm fib}})_L,
\end{align}
where the 2-dimensional operator-valued vectors
\begin{align}
(m_{\mathcal P_{\rm fib}})_1[\sigma_1,\sigma^\prime_1]
&=
\delta_{\sigma_1,0}
\begin{pmatrix}
[\sigma^z]_{\sigma_1,\sigma^\prime_1} & [\mathbbm 1]_{\sigma_1,\sigma^\prime_1}
\end{pmatrix}
\\
(m_{\mathcal P_{\rm fib}})_L[\sigma_L,\sigma^\prime_L]
&=
-\frac{1}{4}\, \delta_{\sigma_L,0}
\begin{pmatrix}
[\mathbbm 1 +\sigma^z]_{\sigma_L,\sigma^\prime_L} & [\sigma^z-3\mathbbm 1]_{\sigma_L,\sigma^\prime_L}
\end{pmatrix}^\mathsf T
\end{align}
and the $2\times 2$ operator-valued matrices
\begin{align}
(M_{\mathcal P_{\rm fib}})_i[\sigma_i,\sigma^\prime_i]
&=
-\frac{1}{4}
\begin{pmatrix}
[\sigma^z+\mathbbm 1]_{\sigma_i,\sigma^\prime_i} & [\sigma^z+\mathbbm 1]_{\sigma_i,\sigma^\prime_i} \\
[\mathbbm 1-3\sigma^z]_{\sigma_i,\sigma^\prime_i} & [\sigma^z-3\mathbbm 1]_{\sigma_i,\sigma^\prime_i}
\end{pmatrix}.
\end{align}
\end{subequations}
Given these expressions we can write
\begin{subequations}
\label{eq: Sn MPS}
\begin{align}
|\mathcal S_n\rangle \propto \sum_{\{\sigma_i\}^{L}_{i=1}}b_1[\sigma_1]B_2[\sigma_2]\dots B_{L-1}[\sigma_{L-1}]b_L[\sigma_{L}]\ |\{\sigma_i\}^{L}_{i=1}\rangle,
\end{align}
where
\begin{align}
b_{1}[\sigma_1]&=\sum_{\tau_1,\sigma^\prime_1}(m_{\mathcal P_{\rm fib}})_1[\sigma_1,\tau_1]\otimes(m_{(\sigma^+_\pi)^n})_1[\tau_1,\sigma^\prime_1]\otimes a_1[\sigma^\prime_1]\\
b_{L}[\sigma_L]&=\sum_{\tau_L,\sigma^\prime_L}(m_{\mathcal P_{\rm fib}})_L[\sigma_L,\tau_L]\otimes(m_{(\sigma^+_\pi)^n})_ L[\tau_L,\sigma^\prime_L]\otimes a_L[\sigma^\prime_L]
\end{align}
and
\begin{align}
B_{i}[\sigma_i]&=\sum_{\tau_i,\sigma^\prime_i}(M_{\mathcal P_{\rm fib}})_i[\sigma_i,\tau_i]\otimes(M_{(\sigma^+_\pi)^n})_ i[\tau_i,\sigma^\prime_i]\otimes A_i[\sigma^\prime_i],
\end{align}
\end{subequations}
which is an MPS with bond dimension $2\chi_n$.

Armed with the MPS expression for $|\mathcal S_n\rangle$ in Eq.~\eqref{eq: Sn MPS}, we can now compute the reduced density matrix $\rho_p$ for a $p$-site block, whose matrix elements are
\begin{align}
\langle\sigma_{i}\dots\sigma_{i+p-1}|\rho_p|\sigma^\prime_{i}\dots\sigma^\prime_{i+p-1}\rangle = \text{tr}\left[B_{i}[\sigma_i]\dots B_{i+p-1}[\sigma_{i+p-1}]\left(B_{i}[\sigma^\prime_i]\dots B_{i+p-1}[\sigma^\prime_{i+p-1}]\right)^\dagger\right],
\end{align}
for any $i$ in the bulk of the chain (note that the choice of $i$ is immaterial due to translation invariance in the bulk).  Our goal is to find a set of local projectors $\mathcal P_i$ such that 1) $\mathcal P_i|\mathcal S_n\rangle=0$ for any $n$ and $i$, and 2) the states $|\mathcal S_n\rangle$ are the \textit{only} states annihilated by all $\mathcal P_i$.  This can be accomplished by computing the \textit{null space} $\text{ker}(\rho_p)$ of the $2^p\times 2^p$ Hermitian matrices $\rho_p$: if $|\psi\rangle$ is in the null space of $\rho_p$, then $\mathcal P_\psi\rho_p=|\psi\rangle\langle\psi|\rho_p=0$.  This implies the existence of a $p$-site operator that annihilates the state $|\mathcal S_n\rangle$.

Carrying this procedure out for $\rho_2$, we find for any $n$ that
\begin{align}
\text{ker}(\rho_2)=\text{span}\left\{|11\rangle\right\}.
\end{align}
This implies that, for any $i$ and $n$,
\begin{align}
\label{eq: PPSn}
P^1_{i}P^1_{i+1}\, |\mathcal S_n\rangle=0,
\end{align}
which is a simple consequence of the fact that the states $|\mathcal S_n\rangle$ obey the Fibonacci constraint.  Thus we have found a set of projectors satisfying criterion 1).  However, there are exponentially many states satisfying the Fibonacci constraint, so we have not yet satisfied criterion 2); we need additional terms to isolate the states $|\mathcal S_n\rangle$. The next nontrivial projection operator comes from considering $\rho_4$, where we find
\begin{align}
\text{ker}(\rho_4)=\text{span}\left\{\frac{|0100\rangle+|0010\rangle}{\sqrt 2}\right\}\cup\text{span}\{\substack{\text{all $4$-site configurations violating}\\ \text{the Fibonacci constraint}}\}.
\end{align}
This implies that in addition to Eq.~\eqref{eq: PPSn} we have, for any $i$ and $n$,
\begin{align}
\label{eq: PHeisPSn}
P^0_{i-1}\left[\frac{1}{4}+\frac{1}{2}(\sigma^+_i\sigma^-_{i+1}+\text{H.c.})-\frac{1}{4}\sigma^z_i\sigma^z_{i+1}\right]P^0_{i+2}\, |\mathcal S_n\rangle = 0.
\end{align}
The projection operators $\mathcal P_i$ defined in Eq.~\eqref{eq: Pi def} are obtained by summing the operators appearing on the left hand sides of Eqs.~\eqref{eq: PPSn} and \eqref{eq: PHeisPSn}.  We have verified by exact numerical diagonalization of Eq.~\eqref{eq: Hu} at system sizes up to $L=22$ that the states $|\mathcal S_n\rangle$ are the only ground states of the Hamiltonian \eqref{eq: Hu}.

\section{MPS form of the states \eqref{eq: RK}}
\label{sec: xi MPS}

We demonstrate here that the initial states defined in Eq.~\eqref{eq: RK} can be written as a family of MPSs with bond dimension 2.  To write these states in MPS form, we first express them as
\begin{align}
|\xi\rangle
\propto
\mathcal P_{\rm fib}
\prod^{i_L}_{i=i_1}\left[1+(-1)^i\, \xi\,\sigma^+_i\right]|\Omega\rangle.
\end{align}
We then write 
\begin{subequations}
\begin{align}
\label{eq: xi MPS 1}
\prod^{i_L}_{i=i_1}\left[1+(-1)^i\, \xi\,\sigma^+_i\right]|\Omega\rangle=\sum_{\{\sigma_i\}^{L}_{i=1}}c_1[\sigma_1]C_2[\sigma_2]\dots C_{L-1}[\sigma_{L-1}]c_L[\sigma_{L}]\ |\{\sigma_i\}^{L}_{i=1}\rangle,
\end{align}
with
\begin{align}
c_{1,L}[\sigma_{1,L}]=\delta_{\sigma_{1,L},0}
\end{align}
for OBC and
\begin{align}
C_{i}[\sigma_i]=(-1)^i\, \xi\, \delta_{\sigma_{i},1}+\delta_{\sigma_i,0}.
\end{align}
\end{subequations}
Next we apply $\mathcal P_{\rm fib}$ to Eq.~\eqref{eq: xi MPS 1} using the MPO expression \eqref{eq: Pfib MPO}, thereby obtaining the MPS
\begin{subequations}
\begin{align}
|\xi\rangle\propto \sum_{\{\sigma_i\}^{L}_{i=1}}d_1[\sigma_1]D_2[\sigma_2]\dots D_{L-1}[\sigma_{L-1}]d_L[\sigma_{L}]\ |\{\sigma_i\}^{L}_{i=1}\rangle,
\end{align}
where
\begin{align}
d_{1}[\sigma_{1}]&=\sum_{\sigma^\prime_1}(m_{\mathcal P_{\rm fib}})_1[\sigma_1,\sigma^\prime_1]\otimes c_{1}[\sigma^\prime_{1}] = \delta_{\sigma_1,0}\,\begin{pmatrix}-1 & 1\end{pmatrix}\\
d_{L}[\sigma_{L}]&=\sum_{\sigma^\prime_L}(m_{\mathcal P_{\rm fib}})_L[\sigma_L,\sigma^\prime_L]\otimes c_{L}[\sigma^\prime_{L}] = \delta_{\sigma_L,0}\,\begin{pmatrix}0 & 1\end{pmatrix}^\mathsf{T}
\end{align}
and
\begin{align}
D_{i}[\sigma_i]&=\sum_{\sigma^\prime_i}(M_{\mathcal P_{\rm fib}})_i[\sigma_i,\sigma^\prime_i]\otimes C_{i}[\sigma^\prime_{i}] =
\delta_{\sigma_i,0}\,
\begin{pmatrix}
0&0\\
-1&1
\end{pmatrix}
+\delta_{\sigma_i,1}\,
(-1)^i\, \xi\,
\frac{1}{2}
\begin{pmatrix}
-1 & -1\\
1 & 1
\end{pmatrix}.
\end{align}
\end{subequations}
Importantly, note that $D_{i}[1]D_{i+1}[1]=0$, ensuring that the Fibonacci constraint holds as it should.

\end{widetext}

\bibliography{refs_zero}

\end{document}